\pdfoutput=1
\documentclass[aps, rmp, twocolumn, groupedaddress, floatfix, nofootinbib]{revtex4-1}

\usepackage{amsmath, amsfonts, amssymb}
\usepackage{graphicx}
\newcommand{\eq}[1]{Eq.~(\ref{#1})}
\newcommand{\Eq}[1]{Equation~(\ref{#1})}
\newcommand{\fig}[1]{Fig.~\ref{#1}}
\newcommand{\Fig}[1]{Figure~\ref{#1}}
\newcommand{\eqs}[2]{Eqs.~(\ref{#1}) and (\ref{#2})}
\newcommand{\Eqs}[2]{Equations~(\ref{#1}) and (\ref{#2})}

\newcommand{\threeeqs}[3]{Eqs.~(\ref{#1}), (\ref{#2}), and (\ref{#3})}

\newcommand{\tbl}[1]{Table~\ref{#1}}

\newcommand{\pfix}{p_\mathrm{fix}}
\newcommand{\Un}{U_n}
\newcommand{\Ub}{U_b}
\newcommand{\rhob}{\rho_b}
\newcommand{\rhoo}{\rho_0}
\newcommand{\Utot}{U_\mathrm{tot}}
\newcommand{\seff}{s_\mathrm{eff}}
\newcommand{\Ueff}{U_\mathrm{eff}}
\newcommand{\Xc}{X_c}
\newcommand{\vo}{v_0}
\newcommand{\Ro}{R_0}
\newcommand{\Xavg}{\overline{X}}
\newcommand{\Mavg}{\overline{M}}
\newcommand{\Mbavg}{\Mavg_b}
\newcommand{\sfixed}{\langle s \rangle_f}
\newcommand{\dx}{\Delta X}
\newcommand{\dxavg}{\Delta \Xavg}
\newcommand{\sigmadx}{\sigma_{\mathrm{err},\dxavg}}
\newcommand{\sigmax}{\sigma_{\mathrm{err},\Xavg}}
\newcommand{\Lb}{L_b}
\newcommand{\tmax}{t_\mathrm{max}}

\begin{document}

\title{The impact of macroscopic epistasis on long-term evolutionary dynamics}
\author{Benjamin H. Good$^{1}$}
\author{Michael M. Desai$^{1}$}
\affiliation{\mbox{${}^1$Department of Organismic and Evolutionary Biology, Department of Physics, and} \mbox{FAS Center for Systems Biology, Harvard University, Cambridge MA 02138}}

\begin{abstract}
Genetic interactions can strongly influence the fitness effects of individual mutations, yet the impact of these epistatic interactions on evolutionary dynamics remains poorly understood. Here we investigate the evolutionary role of epistasis over 50,000 generations in a well-studied laboratory evolution experiment in \emph{E. coli}. The extensive duration of this experiment provides a unique window into the effects of epistasis during long-term adaptation to a constant environment. Guided by analytical results in the weak-mutation limit, we develop a computational framework to assess the compatibility of a given epistatic model with the observed patterns of fitness gain and mutation accumulation through time. We find that a decelerating fitness trajectory alone provides little power to distinguish between competing models, including those that lack any direct epistatic interactions between mutations. However, when combined with the mutation trajectory, these observables place strong constraints on the set of possible models of epistasis, ruling out many existing explanations of the data. Instead, we find that the data are consistent with ``two-epoch'' model of adaptation, in which an initial burst of diminishing returns epistasis is followed by a steady accumulation of mutations under a constant distribution of fitness effects. Our results highlight the need for additional DNA sequencing of these populations, as well as for more sophisticated models of epistasis that are compatible with all of the experimental data.
\end{abstract}

\maketitle

\section*{Introduction}

\noindent A central feature of evolutionary adaptation is that the space of potential innovations can vary with the evolutionary history of a population. Examples are common in the microbial world: the ability to import a nutrient may be beneficial only if a mechanism has evolved to utilize it \citep{quandt:etal:2014}, while a previously advantageous drug resistance mutation can be rendered obsolete by the acquisition of a second resistance allele \citep{weinreich:etal:2006}. This capacity for evolutionary feedback is quantified in terms of \emph{epistasis}, which measures how the effect of a mutation depends on the genetic background in which it arises. In principle, epistasis can lead to widespread historical contingency, and can fundamentally alter the dynamics of adaptation \citep{wright:1932, gould:1989}. But in practice, the long-term evolutionary impact of epistasis remains largely uncharacterized.

Empirical patterns of epistasis are most commonly measured using a direct approach (see \citet{deVisser:krug:2014} for a recent review). Candidate mutations are introduced into a set of genetic backgrounds via crossing or other genetic reconstruction techniques, and the fitnesses of the reconstructed genotypes are measured using competitive fitness assays or related proxies. These data yield a functional relationship between the fitness effect of a mutation and its genetic background, with the traditional pairwise epistasis emerging as a special case when the backgrounds are single mutants. We will use to term \emph{microscopic epistasis} to refer to these measurements, since they track the background-dependence of individual mutations. Of course, the precise background dependence of any given mutation is essentially an empirical matter: the fitness effects depend on the biological details of the organism, its environment, and the identities of the mutations themselves. Empirical estimates of these quantities therefore provide valuable insight into the physiological and biophysical properties of the organism \citep{segre:etal:2005, st-onge:etal:2007, kinney:etal:2010, costanzo:etal:2010}.

With enough reconstructions, one can also obtain information about the larger-scale structure of the fitness landscape. For example, one can assess whether interactions between mutations are generally antagonistic or synergistic \citep{jasnos:korona:2007}, or estimate the prevalence of sign epistasis \citep{weinreich:etal:2006} and overall levels of modularity \citep{segre:etal:2005}. These questions are typically quantified using statistical summaries of the microscopic epistasis (e.g., the distribution of pairwise epistasis values), which are aggregated over a large ensemble of mutations and genetic backgrounds. While the biological interpretation of these statistics is sometimes unclear, they can in principle exhibit regular and generalizable patterns. This makes them potentially amenable to comparison with simple fitness landscape models, such as Fisher's geometrical model \citep{fisher:1930}, the uncorrelated landscape \citep{kingman:1978, kauffman:levin:1987}, and the NK landscape \citep{kauffman:weinberger:1989}. Yet while these statistics are designed to capture global properties of the fitness landscape, they are still fundamentally microscopic in nature, since they can only be estimated from microscopic (i.e. reconstruction-based) measurements. As such, they suffer from the same throughput limitations as any other reconstruction-based method, and one can only focus on a small subset of possible genotypes.

However, an evolving population typically explores a much larger number of genotypes than is feasible to reconstruct experimentally. The evolutionary dynamics depends on the entire \emph{distribution} of fitness effects (``the DFE''), and on how this distribution varies among different genetic backgrounds. We denote this background-dependent DFE by $\rho(s|\vec{g})$: the fraction of mutations with fitness effect $s$ in genetic background $\vec{g}$. In contrast to the statistics above, the background dependence of the DFE is a \emph{macroscopic} form of epistasis, since it includes no information about the background-dependence of any individual mutation. Like many macroscopic quantities, there is not a one-to-one correspondence between the underlying microscopic epistasis and its macroscopic manifestation. For example, one can imagine a scenario where epistasis changes the identities of beneficial mutations after every substitution, but in a way that preserves the overall shape of the DFE. In this case, the widespread patterns of microscopic epistasis are masked at the macroscopic level, and the dynamics of adaptation will be indistinguishable from a non-epistatic scenario. At the opposite extreme, the DFE can change even without any microscopic epistasis once selection starts to deplete the finite supply of beneficial mutations. In this case, the dynamics of adaptation will show signatures of macroscopic epistasis even though there are no direct interactions between mutations.

Despite the potential importance of macroscopic epistasis in evolutionary adaptation, it remains less well-characterized than its microscopic counterpart. In principle, it is possible to measure the background dependence of the DFE directly, by assaying the fitness of large libraries of random mutants \citep{silander:etal:2007, maclean:etal:2010, miller:etal:2011, bank:etal:2014}. However, such studies suffer from similar throughput limitations as the microscopic approach above. These throughput limitations are compounded by the fact that the most important changes in the DFE, from an evolutionary perspective, are often located in difficult-to-sample regions such as the high-fitness tail \citep{good:etal:2012}. 

To avoid these these issues, a number of studies have focused on the evolutionary outcomes themselves, associating observed differences in the adaptability of different strains with differences in the underlying DFE \citep{burch:chao:2000, silander:etal:2007, barrick:etal:2010, woods:etal:2011, kryazhimskiy:etal:2012, perfeito:etal:2014, kryazhimskiy:etal:2014}. In principle, this approach offers the greatest sensitivity for detecting relevant differences in the DFE among related genetic backgrounds. However, it does so by transforming the measurement into an inverse problem: the patterns of macroscopic epistasis must ultimately be inferred from the dynamics of a few observable quantities (e.g. changes in fitness over time or across experimental treatments), which depend on the complex population genetics of an evolving microbial population \citep{lang:etal:2013, frenkel:etal:2014}. Thus, while it is easy to demonstrate the \emph{existence} of macroscopic epistasis with this approach, it is difficult to associate the observed differences in adaptability with the precise changes in the underlying DFE. This, in turn, has made it hard to distinguish between competing models of epistasis when interpreting the results of the experiment \citep{kryazhimskiy:etal:2014, frank:2014}.  

In the present manuscript, we propose a general framework for quantifying patterns of macroscopic epistasis from observed differences in adaptability. We then use this framework to investigate the role of epistasis in a well-studied laboratory evolution experiment in \emph{E. coli} \citep{wiser:etal:2013}. By analyzing the differences in the dynamics of adaptation through time, we can make inferences about the changes in the DFE that have accumulated over the course of the experiment. These changes constitute the most basic form of epistasis that arises during adaptation to a constant environment. Similar to  \citet{kryazhimskiy:etal:2009}, we focus on two simple summaries of the dynamics: the competitive fitness and the total number of genetic changes relative to the ancestor. We use a combination of theory and numerical simulations to investigate how well these data are explained by several popular models of macroscopic epistasis, including the recently proposed diminishing returns model of \citet{wiser:etal:2013}. We find that fitness measurements alone have little power to discriminate between different models of epistasis, while the addition of genetic information is sufficient to rule out many existing explanations of the data. Together, these results highlight the need for more sophisticated models of macroscopic epistasis that are compatible with \emph{all} of the experimental data, as well as additional DNA sequence data to test their predictions.

\section*{Results}

\subsection*{Fitness and mutation trajectories in the LTEE}

\begin{figure*}[t]
\centering
\includegraphics[width=\textwidth]{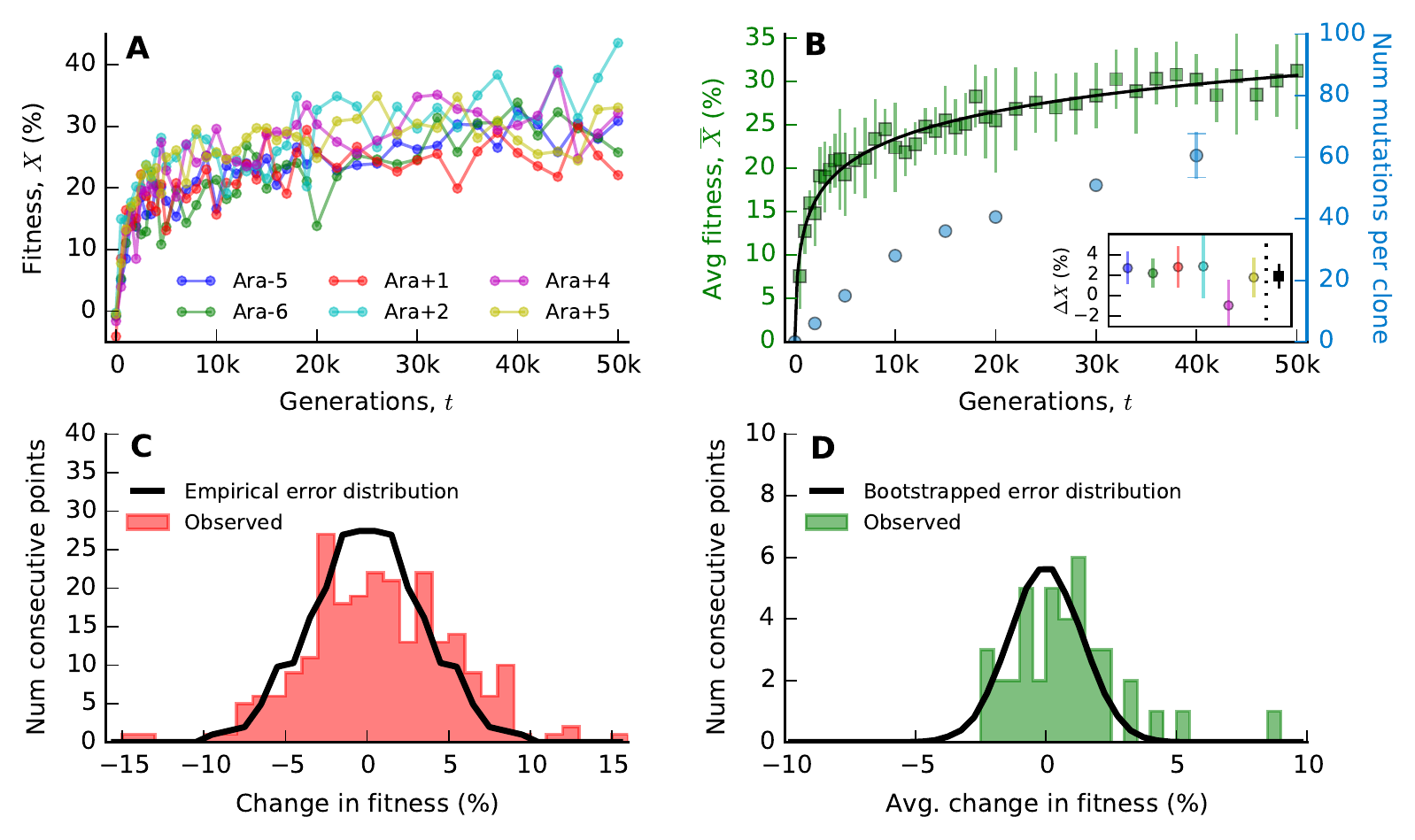}
\caption{Fitness and mutation trajectories in the LTEE. (A) Individual fitness trajectories for the six complete, non-mutator populations analyzed by \citet{wiser:etal:2013}. Each point is the average of two independent competition assays, with fitness estimated from \eq{eq:fitness}. (B) The average fitness trajectory for the six populations in panel A (green squares), with $\pm 2$ stderr confidence intervals. For comparison, the solid line depicts the logarithmic trajectory in \eq{eq:wiser-trajectory}, with estimated parameters $X_c \approx 4.6 \times 10^{-2}$ and $\vo \approx 7.7 \times 10^{-4}$. The blue circles depict the average number of mutations in single clones sampled from the LTEE (see Appendix), with $\pm 2$ stderr confidence intervals for timepoints with more than one sampled population. (Inset) Independent measurements of the change in fitness, $\dx$, between generation 40,000 and 50,000 for the six populations in panel A (left, circles) and their average (right, square), with $\pm 2$ stderr confidence intervals. (C) The change in fitness between consecutive timepoints, pooled across all six populations in panel A. The black curve shows the empirical distribution of measurement errors, defined as half of the difference between replicate fitness measurements. (D) The average change in fitness between consecutive timepoints. The black curve shows the bootstrapped distribution of measurement errors, obtained by repeatedly averaging six randomly chosen errors from the empirical distribution in panel C. The standard deviation of this distribution is $\sigmax \approx 1.4\%$.  \label{fig:trajectory}}
\end{figure*}

\noindent In the long-term evolution experiment (LTEE) conducted by Lenski and collaborators, twelve populations of \emph{Escherichia coli} were founded from a single common ancestor \citep{lenski:etal:1991} and propagated in a constant environment for more than $60,000$ generations (see \citet{wiser:etal:2013} for a recent summary of experimental details). A central observable quantity is the fitness of the evolved populations, which can be measured using competition assays with a marked ancestor. If $f_i$ and $f_f$ denote the frequencies of the evolved strain at the beginning and end of the competition assay, then the (log)-fitness, $X$, is given by
\begin{align}
X \equiv \frac{1}{\Delta t} \log \left[ \frac{f_f}{1-f_f} \frac{1-f_i}{f_i} \right]\, , \label{eq:fitness}
\end{align}
where $\Delta t$ is the duration of the competition in generations. Note that this definition of fitness differs from the traditional measure $W$ reported in previous studies of the LTEE. Although the two measures are correlated, \eq{eq:fitness} provides a more direct connection to the population genetic theory described in the following sections.

Using the fitness assays reported in \citet{wiser:etal:2013}, we calculated the fitness defined by \eq{eq:fitness} for each population at approximately 40 timepoints during the first 50,000 generations of evolution (see Appendix). We plot the fitness trajectories for the six complete, non-mutator populations in \fig{fig:trajectory}A. Measurement error estimated from replicate assays is substantial ($\mathrm{stderr} \sim 3\%$, \fig{fig:trajectory}C), leaving us with little power to distinguish fluctuations in individual trajectories. Instead, we pool all six populations and focus on the \emph{average fitness trajectory} $\Xavg(t)$ (\fig{fig:trajectory}B). Bootstrap resampling from the errors in \fig{fig:trajectory}C suggests that the measurement error in $\Xavg(t)$ is smaller ($\mathrm{stderr} \sim 1\%$) and more normally distributed (\fig{fig:trajectory}D). However, even for the average fitness trajectory, the fluctuations between neighboring timepoints still fall within experimental uncertainty, so we can only obtain robust inferences from long-term trends in the data. 

The most striking trend is the pronounced slowdown in the rate of adaptation during the course of the experiment: nearly two-thirds of the total fitness was gained in the first 5,000 generations of evolution. This deceleration is inconsistent with a constant DFE, which would predict that the average fitness increases linearly with time. Instead, the slowdown in the rate of adaptation has long been interpreted as a signature of \emph{diminishing returns epistasis}, consistent with the approach to a fitness plateau \cite{lenski:travisano:1994}. Previous work has argued that the shape of the deceleration is best captured by a logarithmic fitness trajectory, 
\begin{align} 
\Xavg(t) = \Xc \log \left( 1 + \frac{\vo t}{\Xc} \right) \, , \label{eq:wiser-trajectory} 
\end{align} 
where $\vo$ gives the initial rate of fitness increase and $\Xc$ controls the severity of the slowdown \citep{sibani:etal:1998, kryazhimskiy:etal:2009, wiser:etal:2013}; the best-fit parameters are shown in \fig{fig:trajectory}B. The shape of this trajectory,  in combination with more precise measurements of the change in fitness between generations 40,000 and 50,000 ($\Delta X$, see inset of \fig{fig:trajectory}B), have been used to argue that fitness is still increasing in the LTEE \citep{wiser:etal:2013}, rather than asymptoting to a fitness peak \citep{lenski:travisano:1994}. 

More recent work has tried to use the shape of the fitness trajectory to make inferences about the underlying model of epistasis in the LTEE \citep{kryazhimskiy:etal:2009, wiser:etal:2013, frank:2014}. However, to truly distinguish between different models, we must move beyond the simple curve fitting implied by \eq{eq:wiser-trajectory} and postulate a set of concrete population genetic models that can be used to generate predictions for $\overline{X}(t)$. The likelihood of the observed fitness trajectory can then be written in the form
\begin{align}
p(\Xavg_\mathrm{obs} | \theta) = \int p_\mathrm{err}(\Xavg_\mathrm{obs}-\Xavg) \cdot p_\mathrm{evol}(\Xavg|\theta) \, d\Xavg \, . \label{eq:likelihood}
\end{align}
Here, $p_\mathrm{evol}(\Xavg|\theta)$ is the probability distribution of the data vector $(\Xavg(t_0),\ldots \Xavg(t_n),\Delta X)$ in the underlying model, which depends on some set of parameters $\theta$, and $p_\mathrm{err}(\vec{\epsilon})$ is the distribution of measurement errors, which we assume to be independent and normally distributed with variance $\sigmax \approx 1.4\%$ for each timepoint of $\Xavg(t)$ (\fig{fig:trajectory}) and $\sigmadx \approx 0.4\%$ for $\dxavg$ (\fig{fig:trajectory}B, inset). By computing this likelihood, we can assess the fit of a given model using standard statistical techniques (see Appendix). In contrast to the curve-fitting approach of earlier work, this method correctly accounts for inherent stochasticity of the evolutionary process, which can lead to correlated fluctuations in the observed fitness trajectory. Yet in practice, it is often difficult to compute the likelihood in \eq{eq:likelihood} because the model distribution $p_\mathrm{evol}(\Xavg|\theta)$ is unknown. This is largely due to the large population size of the LTEE ($N \approx 3 \times 10^7$), which makes it difficult to analyze even the simplest population genetic models \citep{desai:fisher:2007}. To avoid these issues, we use computer simulations of the model to obtain accurate predictions of the fitness trajectory (see Appendix), computing the approximate likelihood function as
\begin{align}
p(\Xavg_\mathrm{obs} | \theta) \approx \frac{1}{n} \sum_{i=1}^n  p_\mathrm{err} \left( \Xavg_\mathrm{obs}-\Xavg_{\mathrm{sim}(\theta),i} \right) . \label{eq:approx-likelihood}
\end{align}

Unfortunately, regardless of the method used for inference, we will demonstrate that there is little power to distinguish between different models of epistasis based on the fitness trajectory alone. As noted by \citet{frank:2014}, it is relatively easy to devise an epistatic model that reproduces the observed fitness trajectory in \fig{fig:trajectory}B, and we outline several specific examples below. Fortunately, the average fitness trajectory is not the only quantity that has been measured in the LTEE. DNA sequences from a small number of clones are available for several timepoints in a subset of the lines \citep{barrick:etal:2009, wielgoss:etal:2011, wielgoss:etal:2013}. Although this genetic data is more sparse than the fitness measurements, it provides a crucial window into the the molecular changes responsible for the observed patterns of fitness evolution. In \fig{fig:trajectory}B, we plot the average number of genetic differences between the ancestor and a set of clones sampled from the non-mutator populations (see Appendix). When viewed as a function of time, this \emph{mutational trajectory} $\Mavg(t)$ is the natural genetic analogue of the average fitness trajectory $\Xavg(t)$. Any evolutionary model which purports to explain the long term trends in $\Xavg(t)$ must also be consistent with the observed values of $\Mavg(t)$. As we will see below, this turns out to be much more informative than fitting the fitness trajectory on its own.

The most striking feature of the mutation trajectory in \fig{fig:trajectory} is the sheer number of mutations that have accumulated during the experiment. Although the full data no longer support the constant substitution rate observed in the first 10,000 generations of evolution \citep{barrick:etal:2009}, the number of mutations in the later portion of the experiment is still much higher than one might expect based on the fitness trajectory. Roughly half of all mutations accumulated \emph{after} the first 10,000 generations, when rate of fitness increase had already slowed substantially. Of course, some unknown fraction of these mutations are likely to be selectively neutral, as these accumulate continuously at the neutral mutation rate $U_n$ \citep{birky:walsh:1988}. There are no \emph{a priori} estimates of $U_n$, but evidence from the synonymous substitution rate and mutation accumulation lines suggest that a reasonable upper bound is $\Utot \approx 7 \times 10^{-4}$ (File S1). With this estimate, fewer than 30 neutral mutations should have accumulated by generation 40,000, which suggests that most of the $\sim 60$ observed mutations in \fig{fig:trajectory} are beneficial. In fact, the substitution rate in the first 10,000 generations is so rapid that many of these beneficial mutations must be segregating in the population at the same time. Given that the typical fitness effect of a fixed mutation is at most about $10\%$ \citep{khan:etal:2011}, the fixation time of a successful mutation is much longer than the maximum possible waiting time between mutations. As a result, these mutations must compete for fixation within the population --- a process known as \emph{clonal interference} \citep{gerrish:lenski:1998, desai:fisher:2007}. This will prove to be an important factor in the theoretical analysis below.

\subsection*{Macroscopic epistasis from a finite genome}

\noindent Although a decelerating fitness trajectory is a clear signature of \emph{macroscopic} epistasis (i.e., a changing DFE), this does not necessarily imply that \emph{microscopic} epistasis must be at work. The DFE can change even in the absence of epistasis provided that the length of the genome is finite. Given enough time, the population will eventually exhaust the supply of beneficial mutations, and the rate of adaptation will slow substantially. Thus, this non-epistatic scenario offers one of the simplest possible explanations for the decelerating rate of adaptation in the LTEE, provided that it can also \emph{quantitatively} reproduce the trajectories in \fig{fig:trajectory}. 

In the simplest version of this model, the beneficial DFE evolves according to the mean-field dynamics,
\begin{align} 
\Lb \partial_t \rhob(s,t) = - N \Ub \rhob(s,t) \pfix(s) \, , \label{eq:dfe-evolution} 
\end{align} 
where $\Lb$ is the number of sites and $\pfix(s)$ is the fixation probability of a new mutation. \Eq{eq:dfe-evolution} accounts for the fact that, once a beneficial mutation fixes, a second mutation at that site is not likely to be beneficial, effectively removing this site from the beneficial portion of the DFE. The overall normalization of $\rhob(s,t)$ will therefore decrease as more mutations are driven to fixation. The rate of change of the DFE in \eq{eq:dfe-evolution} is inversely proportional to $\Lb$, and it vanishes in the limit that $\Lb \to \infty$ as expected. In a true ``finite sites'' model, each of the $\Lb$ beneficial mutations corresponds to a single site in the genome, and the ratio $\Ub/\Lb$ is set by the per-site mutation rate $\mu$. However, \eq{eq:dfe-evolution} also describes the evolution of the DFE in a generalized ``running out of mutations'' model --- for example, there could be $\Lb$ genes which are beneficial to knock out, or $\Lb$ modules to improve \citep{tenaillon:etal:2012, kryazhimskiy:etal:2014}. In these cases, $\Lb$ represents the total number of non-redundant mutations, e.g. the number of genes to knock out, and $\Ub/\Lb$ is the target size of each module. Note that this model assumes that all modules share the same target size; the variable target size case is treated in more detail in File~S1. 

Given a solution for the time-dependent DFE in \eq{eq:dfe-evolution}, the expected fitness and mutation trajectories are given by
\begin{subequations} 
\label{eq:trajectory-equations} 
\begin{align} 
& \partial_t \Xavg(t) = \int s  N \Ub \rhob(s,t) \pfix(s) \, ds , \\ 
& \partial_t \Mbavg(t) = \int N \Ub \rhob(s,t) \pfix(s)\, ds \\
& \partial_t \Mavg(t) = \partial_t \Mbavg(t) + \Un
\end{align} 
\end{subequations} 
Unfortunately, both \Eqs{eq:dfe-evolution}{eq:trajectory-equations} are difficult to solve in general, since the fixation probability also depends on the DFE \citep{good:etal:2012}. Despite this difficulty, we can gain considerable qualitative insight by focusing on the strong-selection, weak-mutation (SSWM) limit, where the fixation probability is given by Haldane's formula, $\pfix(s) \approx 2 s$ \citep{haldane:1927}. In this limit, the evolution of the DFE greatly simplifies, and the distribution of beneficial fitness effects is given by 
\begin{align} 
\rhob(s,t) = \rhoo(s) e^{- 2 N \Ub s t / \Lb } \, , \label{eq:sswm-dfe-evolution} 
\end{align} 
where $\rhoo(s)$ is the DFE in the ancestral background. The average fitness and mutation trajectories can then be obtained by substituting \eq{eq:sswm-dfe-evolution} into \eq{eq:trajectory-equations} and evaluating the resulting integral. For example, arguments from extreme value theory suggest that the ancestral DFE may often be exponential \citep{gillespie:1984, orr:2002}, which leads to an average fitness trajectory of the form
\begin{align}
\Xavg(t) = \Xc \left[ 1 - \left( 1 + \frac{\vo t}{2 \Xc} \right)^{-2} \right] \, , \label{eq:finite-exponential-trajectory}
\end{align} 
where $\vo = 2 N \Ub \int s^2 \rhoo(s) \, ds$ and $\Xc = \Lb \int s \rhoo(s) \, ds$. However, while this trajectory shares the same qualitative deceleration as the data in \fig{fig:trajectory}, it predicts a much sharper deceleration in the adaptation rate than is actually observed (\fig{fig:trajectory-curvefits}). This shows that for a fixed DFE shape, we will not always be able to quantitatively reproduce the observed fitness trajectory with our finite-sites model. 

However, the situation changes if we are allowed to arbitrarily tune the shape of the DFE to match to the observed fitness trajectory. In particular, we find that the adaptation rate for the DFE in \eq{eq:sswm-dfe-evolution} is proportional to the Laplace transform of $s^{-2} \rhoo(s)$, which leads to an inverse relation of the form
\begin{align} 
\rhoo(s) = \frac{1}{\Lb s^2} \mathcal{L}^{-1} \left\{ \partial_t \Xavg(t) \right\}\left(\frac{2 N \Ub s}{\Lb} \right) \, . \label{eq:general-finite-dfe} 
\end{align} 
In other words, we can reproduce a particular fitness trajectory within our finite sites model by choosing the ancestral DFE to match the expression above. Note that \eq{eq:general-finite-dfe} implicitly assumes that the inverse Laplace transform exists and yields a proper probability distribution. This places certain constraints on the fitness trajectories that we can reproduce with this model, e.g., requiring that $\partial_t \Xavg(t)$ is monotonically decreasing. The intuitive reason for this restriction is clear from the definition of the model: exhausting the supply of beneficial mutations can never lead to an increasing adaptation rate, no matter how exotic the ancestral DFE. Note, however, that deleterious mutations \citep{mccandlish:epstein:plotkin:2014,  mccandlish:otwinowski:plotkin:2014}, clonal interference \cite{desai:fisher:2007}, and the fixation of mutator phenotypes \citep{wiser:etal:2013} can complicate this picture considerably. The other apparent limitation of this model is that the fitness trajectory must be bounded, since the maximum possible fitness that can be attained is $\Xavg(\infty) = \Lb \int_0^\infty s \rhoo(s) \, ds$. At first glance, this would seem to preclude the logarithmic trajectory in \eq{eq:wiser-trajectory}, which has no maximum value. However, since experimental trajectories are only observed over a finite time window, $0 \leq t \leq \tmax$, we can always satisfy this restriction in practice by assuming that $\Xavg(\tmax) \ll \Xavg(\infty)$. For example, the logarithmic fitness trajectory in \eq{eq:wiser-trajectory} corresponds to an ancestral DFE of the form 
\begin{align} \rhoo(s) \propto \left\{ \begin{array}{ll} s^{-2} e^{-s/\sigma} & s > \sigma \epsilon, \, \\ 0 & s \leq \sigma \epsilon , \end{array} \right. \, \label{eq:wiser-dfe} 
\end{align}
where $\epsilon \ll \Xc/\vo t_\mathrm{max}$ is a lower cutoff chosen to maintain normalization (see File S1). The fitting parameters in \eq{eq:wiser-trajectory} are given by $\Xc = \Lb \sigma \epsilon$ and $\vo = 2 N \Ub \epsilon \sigma^2$. 

A similar argument shows that we can also reproduce a given mutation trajectory (subject to the same technical constraints), provided that the ancestral DFE satisfies
\begin{align}
\rhoo(s) = \frac{1}{\Lb s} \mathcal{L}^{-1} \left\{ \partial_t \Mbavg(t) \right\}\left(\frac{2 N \Ub s}{\Lb} \right) \, . \label{eq:other-general-finite-dfe}
\end{align}
However, while we can fit a broad class of fitness and mutation trajectories by choosing the appropriate ancestral DFE, we do not have complete freedom to fit both quantities at the same time. In the weak mutation limit, the average fitness and mutation trajectories in our finite sites model are related by
\begin{align} 
\Mbavg(t) = 2 N \left( \frac{\Ub}{\Lb} \right) \int_0^t \left[ \overline{X}(\infty) - \overline{X}(\tau) \right] \, d\tau \, , \label{eq:finite-mutation-trajectory} 
\end{align}
regardless of the choice of ancestral DFE. By choosing $\rhoo(s)$ to fit the fitness trajectory, we severely constrain the shape of the mutation trajectory (and vice versa), with only an overall scale $N\Ub/L$ that can be tuned to fit the data. For example, the logarithmic fitness trajectory in \eq{eq:wiser-trajectory} implies a constant substitution rate 
\begin{align} 
\Mbavg(t) \approx \frac{\vo t }{2 N \langle s \rangle_f \left( \frac{\Ub}{\Lb} \right)} \, , \label{eq:finite-wiser-mutation-trajectory} \end{align} 
where $\langle s \rangle_f \approx \sigma / \log \left( 1/\epsilon \right)$. This linear increase is inconsistent with the mutation trajectory in \fig{fig:trajectory}, which starts to show deviations from linearity after generation $10,000$.

However, a potential caveat with this analysis is that the mutation trajectory in \eq{eq:finite-wiser-mutation-trajectory} (and much of the analysis preceding it) depends on our assumption of the weak-mutation limit, which requires that $N\Ub \ll 1$. This is often not self-consistent: in the LTEE, the weak-mutation analysis typically leads us to infer parameter values that violate the weak-mutation assumptions. For example, in the finite-sites model defined by \eq{eq:wiser-dfe}, the fitted values of $\Xc$ and $\vo$ in \fig{fig:trajectory} require that $N\Ub \geq 3$, which violates the weak-mutation condition used to derive \eqs{eq:wiser-dfe}{eq:finite-wiser-mutation-trajectory}. Thus, we must turn to our computational framework to rigorously compare this model with the data. 

\begin{figure}[t]
\centering
\includegraphics[width=\columnwidth]{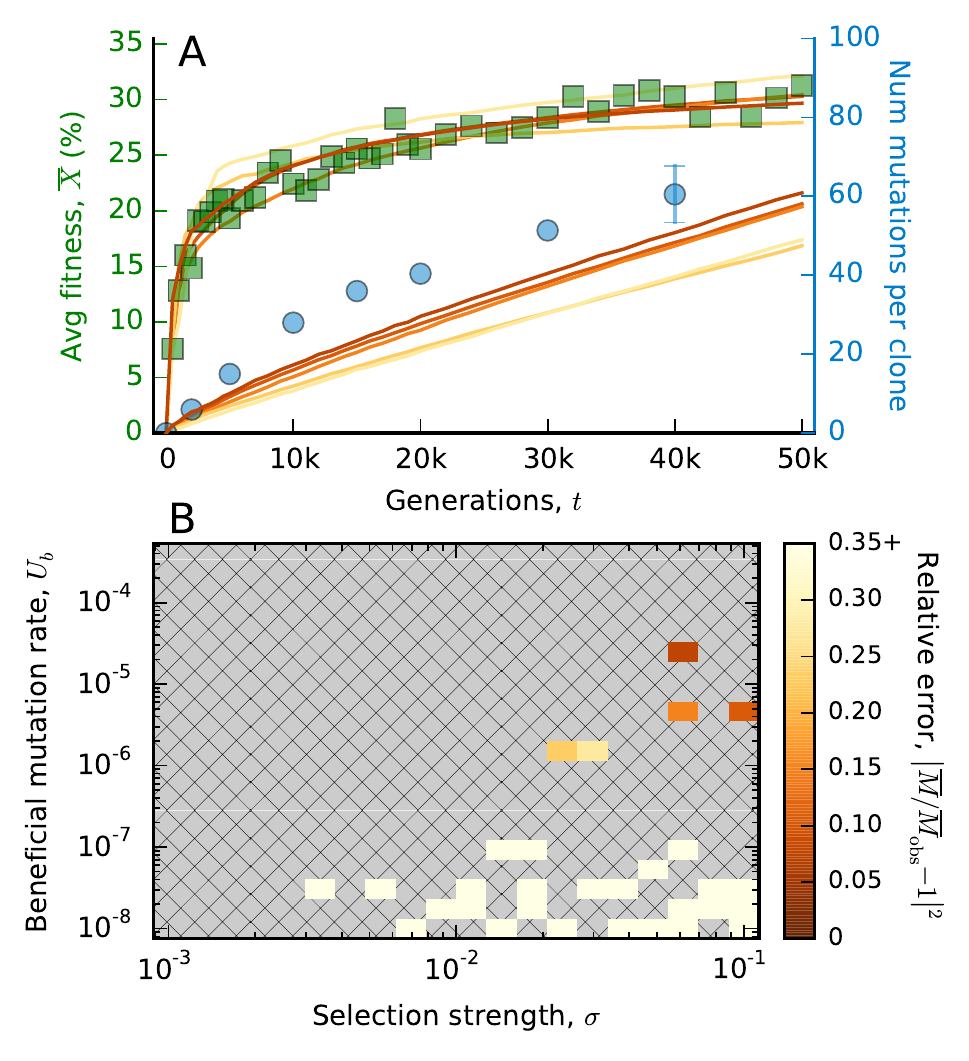} 
\caption{Fitting a finite sites model to the LTEE. (A) Simulated fitness and mutation trajectories for the ancestral DFE in \eq{eq:wiser-dfe} with $\epsilon = 3 \times 10^{-4}$ (solid lines). We have included all simulated combinations of $U$, $\sigma$, and $\Lb$ with $p > 0.05$ and $|\Mavg/\Mavg_\mathrm{obs}-1|^2 < 0.35$. Each line is colored according to the relative error of the mutation trajectory, $|\Mavg/\Mavg_\mathrm{obs}-1|^2$, after fitting the best-fit neutral mutation rate $0 < \Un < \Utot$ by least-squares. For comparison, we have also included the observed fitness and mutation trajectories from \fig{fig:trajectory}B. (B) The relative error of the mutation trajectory for all simulated parameter combinations. For each combination of $U$ and $\sigma$, we plot the minimum error across all simulated values of $\Lb$, and we have only included parameter combinations for which $p > 0.05$. \label{fig:finite-sites-fit}}
\end{figure}

To do so, we performed a grid search over combinations of $\Ub$, $\Lb$, and $\sigma$ for the ancestral DFE defined by \eq{eq:wiser-dfe}. The posterior predictive $p$-value for the fitness trajectory is $p \approx 0.9$ ($\chi^2$ test, see Appendix), which shows that the finite-sites model can still reproduce the observed fitness trajectory in the presence of clonal interference. \Fig{fig:finite-sites-fit} shows the average fitness and mutation trajectories for all parameters with $p > 0.05$. The mutation trajectories also include a best-fit rate of neutral mutations ($0 \leq \Un \leq \Utot-\Ub$) which is fit to minimize the mean squared error from the observed mutation trajectory. Even with this correction, the mutation trajectories remain inconsistent with the data, which allows us to reject the simple finite sites model in \eq{eq:dfe-evolution}. 

\subsection*{Fitness trajectories on an uncorrelated fitness landscape}

\noindent We next consider an alternative model of macroscopic epistasis --- the \emph{uncorrelated fitness landscape} --- which represents the opposite limit of the additive finite-genome models above \citep{kingman:1978, kauffman:levin:1987, gillespie:1984, orr:2002}. In this model, the fitness of every genotype is drawn independently from the same distribution $f(X)$. In our notation, this implies that the DFE is given by 
\begin{align} 
\rho(s|\vec{g}) = f(X(\vec{g})+s) \, , \label{eq:uncorrelated-epistasis} 
\end{align} 
where $X(\vec{g})$ denotes the fitness of genotype $\vec{g}$. This uncorrelated landscape contains extensive microscopic epistasis, with the standard deviation of the pairwise epistasis $\epsilon_{ij} = s_{ij} - s_i - s_j$ on the same order as $s_i$. The fitness effect of a given mutation is therefore barely heritable. However, much of this idiosyncratic microscopic epistasis averages out at the level of the DFE, which depends on the genetic background only through the fitness $X(\vec{g})$.

The dynamics of adaptation become particularly simple when $f(X)$ is exponentially distributed, since the beneficial portion of the DFE remains exponential (with the same mean) regardless of the fitness. Instead, epistasis primarily acts to reduce the beneficial mutation rate via $U_b(X) = U_b e^{-X/\sigma}$, where $\sigma$ is the average fitness effect in the ancestral background. In the weak-mutation limit, this diminishing mutation rate leads to the same logarithmic fitness trajectory as \eq{eq:wiser-trajectory}, with $\Xc = \sigma$ and $\vo = 4 N U_b \sigma^2$ \citep{kryazhimskiy:etal:2009}. Thus, the fitnesses in \fig{fig:trajectory} can also be reproduced in this model of extreme epistasis, in addition to the purely additive model in \eq{eq:wiser-dfe}. However, the corresponding mutation trajectory, 
\begin{align} \Mbavg(t) & = \log \left( 1 + \frac{\vo t}{\Xc} \right) \, , \label{eq:uncorrelated-mutation-trajectory} 
\end{align} 
contains no free parameters. This form of $\Mbavg(t)$ implies a beneficial substitution rate of essentially zero after $t \sim \Xc/\vo$ generations, which is clearly inconsistent with the data, both on a curve-fitting level (\fig{fig:mutation-curvefits}) and in simulation (\fig{fig:uncorrelated-fit}). Thus, while the fitness trajectory is consistent with an uncorrelated landscape, this model is again unable to reproduce the observed mutation trajectory. 

\subsection*{Global fitness-mediated epistasis}

\noindent The general patterns of macroscopic epistasis in the uncorrelated landscape can also be realized in other models which have much less microscopic epistasis. For example, a key simplifying assumption of the uncorrelated landscape is that the effective beneficial mutation rate only depends on the fitness of the genetic background and not on its specific genotype. This leads us to consider a broader class of models of the form 
\begin{align} 
\rho(s|\vec{g}) = \rho(s|X(\vec{g})) \, , 
\end{align}
where the shape of the DFE is similarly mediated by fitness. This form of epistasis has been implicated in recent genetic reconstruction studies \citep{khan:etal:2011, chou:etal:2011, kryazhimskiy:etal:2014}, and it has been hypothesized to describe the patterns of epistasis in the LTEE as well \citep{kryazhimskiy:etal:2009, wiser:etal:2013}. Most of these studies have focused on an even simpler class of models of the form 
\begin{align} 
\rho(s|X) = f(X)^{-1} \rhoo( s / f(X) ) \, , \label{eq:macroscopic-rescaling-epistasis} 
\end{align} 
where the fitness-dependence of the DFE is given by a simple change of scale. We assume by convention that $f(0) = 1$, so that $\rhoo(s)$ represents the ancestral DFE. In the weak mutation limit, the fitness trajectories for \eq{eq:macroscopic-rescaling-epistasis} must satisfy the implicit relation
\begin{align}
t(\Xavg) = \frac{1}{\vo} \int_0^{\Xavg} \frac{dX}{f(X)^2} \, , \label{eq:macroscopic-rescaling-trajectory}
\end{align}  
where $\vo = 2 N \Ub \int_0^\infty s^2 \rhoo(s) \, ds = \partial_t \Xavg(t) |_{t=0}$. We can then invert this equation to solve for $f(X)$ as a function of the fitness trajectory:
\begin{align}
f(X) = \left[ \frac{ \partial_t \Xavg(t) |_{t = \Xavg^{-1}(X)} }{\partial_t \Xavg(t) |_{t = 0}} \right]^{1/2} \, . \label{eq:general-fitness-mediated-epistasis}
\end{align}
Thus, like the finite-sites model above, we can reproduce a given fitness trajectory with the rescaled DFE in \eq{eq:macroscopic-rescaling-epistasis} by choosing the correct form for $f(X)$. Note, however, that \eq{eq:general-fitness-mediated-epistasis} implicitly assumes that the right-hand side exists and is a real-valued function, which is satisfied for all $\partial_t \Xavg(t) > 0$. This is a less restrictive condition than we found for the finite sites model in \eq{eq:general-finite-dfe}, which reflects the fact that fitness-mediated epistasis can generate accelerating as well as decelerating fitness trajectories with the appropriate choice of $f(X)$. 

We can realize this model microscopically by assuming that fitness effects of individual mutations obey the same scaling relation, \begin{align} 
s(X) = s_0 f(X) \, , \label{eq:microscopic-rescaling-epistasis} 
\end{align} 
which allows us to make predictions for microscopic quantities like the fitness effects of reconstructed strains. However, there is not complete freedom to choose $f(X)$ in this microscopic model, since the combined effects of a sequence of mutations must commute with each other. The only rescaling that satisfies this commutative property is the linear relation $f(X) = 1 - X/\Xc$, where $\Xc$ represents the global fitness maximum. In this case, the fitness effect of each mutation is scaled by the fractional distance to the peak, similar to the ``stick-breaking'' model of \citet{nagel:etal:2012}. In the weak-mutation limit, this model reproduces the hyperbolic fitness trajectory
\begin{align}
\Xavg(t) = \vo t \left( 1 + \frac{\vo t}{\Xc} \right)^{-1} \, , \label{eq:hyperbolic-trajectory}
\end{align}
which has been used to fit the LTEE fitness data in previous studies \citep{lenski:travisano:1994}. However, as shown by \citet{wiser:etal:2013}, \eq{eq:hyperbolic-trajectory} provides a relatively poor fit to the observed fitness trajectory (\fig{fig:trajectory-curvefits}), even after accounting for clonal interference (posterior predictive $p < 10^{-3}$). This allows us to rule out all microscopic models of the form in \eq{eq:microscopic-rescaling-epistasis}.

\begin{figure}[t]
\centering
\includegraphics[width=\columnwidth]{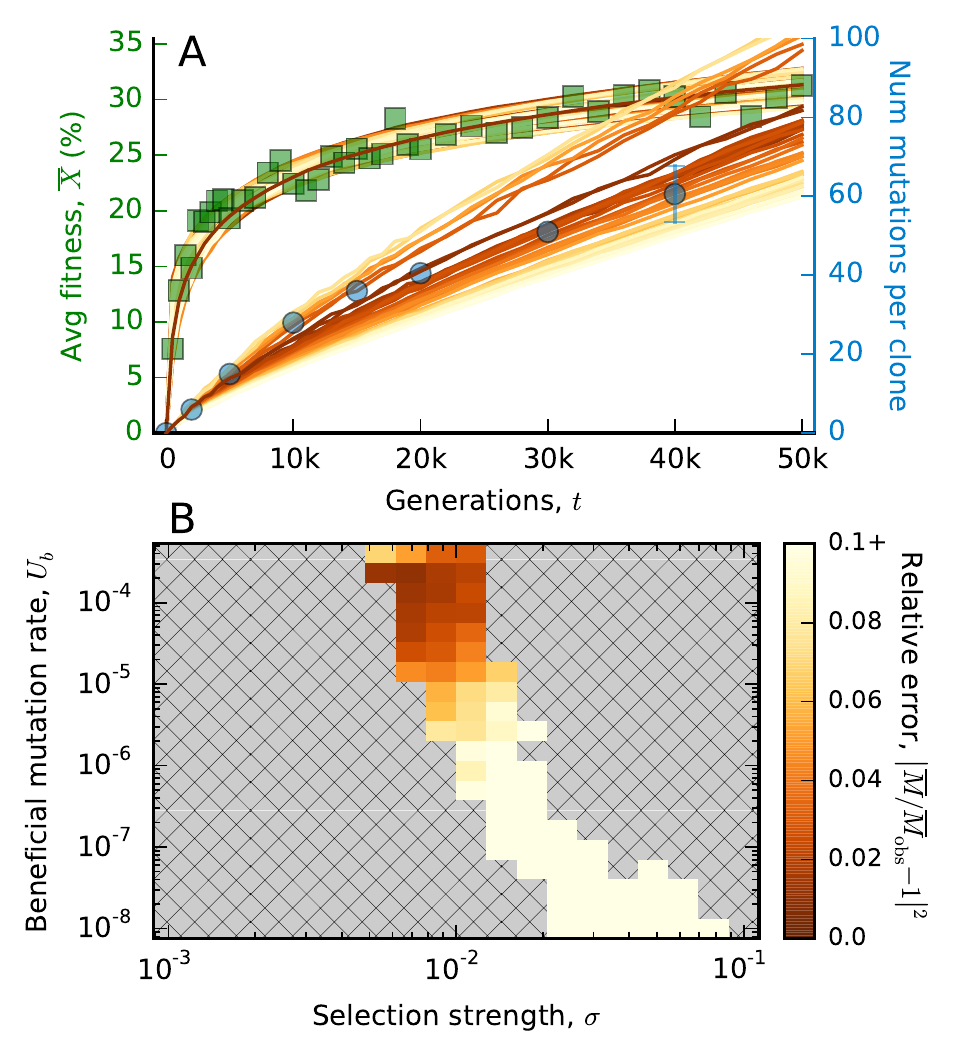} 
\caption{Fitting a fitness-mediated epistasis model to the LTEE data. An analogous version of \fig{fig:finite-sites-fit} constructed for the global diminishing returns model in \eq{eq:wiser-epistasis} with an exponential ancestral DFE. Note the change in scale for the relative error in the mutation trajectory. \label{fig:wiser-fit}}
\end{figure}

For other choices of $f(X)$, \eq{eq:macroscopic-rescaling-epistasis} will hold only in a statistical sense, with a more complicated pattern of microscopic epistasis than predicted by \eq{eq:microscopic-rescaling-epistasis} (File~S1). \citet{wiser:etal:2013} have shown that the logarithmic fitness trajectory in \eq{eq:wiser-trajectory} can be recovered by setting 
\begin{align} 
f(X) = e^{-X/2 \Xc} \, . \label{eq:wiser-epistasis} 
\end{align} 
However, like the additive and uncorrelated models above, the mutation trajectory in this case is strongly constrained by $\Xavg(t)$. In the weak mutation limit, the mutation trajectories for \eq{eq:macroscopic-rescaling-epistasis} must satsify
\begin{align}
\Mbavg(t) = \frac{1}{\sfixed} \int_0^t \left[ \partial_t \Xavg(t)|_{t=0} \cdot \partial_t \Xavg(t) \right]^{1/2} \, dt \, , \label{eq:general-wiser-mutation-trajectory}
\end{align}
with only an overall scale $\sfixed = \int s^2 \rho_0(s) \, ds / \int s \rho_0(s) \, ds$ that can be tuned to fit the data. Note that the shape of $\rhoo(s)$ remains largely unconstrained by this choice: two beneficial DFEs with the same $\sfixed$ can reproduce the same pair of fitness and mutation trajectories, although the required value of $\Ub$ will be different. 

For the logarithmic fitness trajectory in \eq{eq:wiser-trajectory}, \citet{wiser:etal:2013} have shown that the corresponding mutation trajectory grows as a square root of time: 
\begin{align} 
\Mbavg(t) = \frac{2 \Xc}{\sfixed} \left( \sqrt{ 1 + \frac{\vo t}{\Xc} } - 1 \right) \, . \label{eq:wiser-mutation-trajectory}
\end{align} 
At first glance, \eq{eq:wiser-mutation-trajectory} appears to give a decent fit to the observed mutation trajectory (\fig{fig:mutation-curvefits}), although it systematically overestimates the curvature. However, the best-fit scale $\sfixed \sim 5\%$ lies in the clonal interference regime, so we must again turn to our computational framework to rigorously compare this model with the data.

To do so, we performed a grid search over combinations of $U$, $\sigma$, and $X_c$ for an exponential ancestral DFE, $\rho_0(s) \propto \exp(-s/\sigma)$, which was the specific generative model proposed by \citet{wiser:etal:2013}. The posterior predictive $p$-value for the fitness trajectory is $p \approx 0.9$, which shows that the global diminishing returns model can still reproduce the observed fitness trajectory even when $N\Ub > 1$. \Fig{fig:wiser-fit}A shows the average fitness and mutation trajectories for all parameters with $p > 0.05$. As expected, there is a large ``ridge'' of parameter values that reproduce the observed fitness trajectory, but the vast majority of these parameter combinations are inconsistent with the observed mutation trajectory. Those parameters with the best estimates of $\Mavg(t)$ still display some small systematic errors, underestimating the number of mutations in the first part of the experiment and overestimating them later (\fig{fig:wiser-fit}B). The numerical values of these parameters are also wildly unrealistic, since they predict that the mutation rate to fitness effects above $1\%$ is more than a quarter of the genomic point mutation rate. In light of this information, we conclude that the mutation data is inconsistent with the particular global diminishing returns model proposed by \citet{wiser:etal:2013}.

However, our findings in the SSWM limit above suggest that the precise value of the estimated beneficial mutation rate can dramatically vary with the shape of the ancestral DFE, while the predictions of \eqs{eq:general-fitness-mediated-epistasis}{eq:general-wiser-mutation-trajectory} are insensitive to this choice. Similar insensitivity to the DFE has been noted in the clonal interference regime as well \citep{hegreness:etal:2006, desai:fisher:2007, fogle:etal:2008, good:etal:2012, fisher:2013}. In accordance with this intuition, the inferred parameters become more realistic if we truncate the exponential distribution at $s_\mathrm{max} = 4 \sigma$, although the systematic errors in the mutation trajectory remain (\fig{fig:modified-wiser-fit}). Compared to the finite-sites model and uncorrelated landscape above (as well as the original Wiser \emph{et al} model), this modified version of the global diminishing returns model is the only one that can plausibly reproduce \emph{all} of the observed data. 

\subsection*{Evidence for two evolutionary epochs}

\noindent While the discrepancies in the mutation trajectory are too small to reject the fitness-mediated epistasis model outright, these systematic errors still suggest that the model defined by \eq{eq:wiser-epistasis} may be missing a key feature of the experiment. This argues for a degree of caution in interpreting the parameters inferred in \fig{fig:wiser-fit}, particularly far into the future where the errors in the fitness and mutation trajectories start to grow larger. Of course, we could continue to postulate more elaborate models of epistasis to account for the mutation trajectory, and with enough additional parameters this approach is likely to be successful. For example, the fitness and mutation trajectories can both be reproduced by generalized finite sites models where we can tune the individual target sizes (File S1), or by fitness-mediated epistasis models where the overall mutation rate also depends on the fitness. But without a biological basis for choosing among the space of possible models, these additional assumptions are likely to overfit the mutation trajectory and lead to incorrect predictions for other observables (e.g., genetic diversity or variation among lines). Instead, we focus on an alternative class of models that are simpler in some ways, though more complex in others. 

\begin{figure}[t]
\centering
\includegraphics[width=\columnwidth]{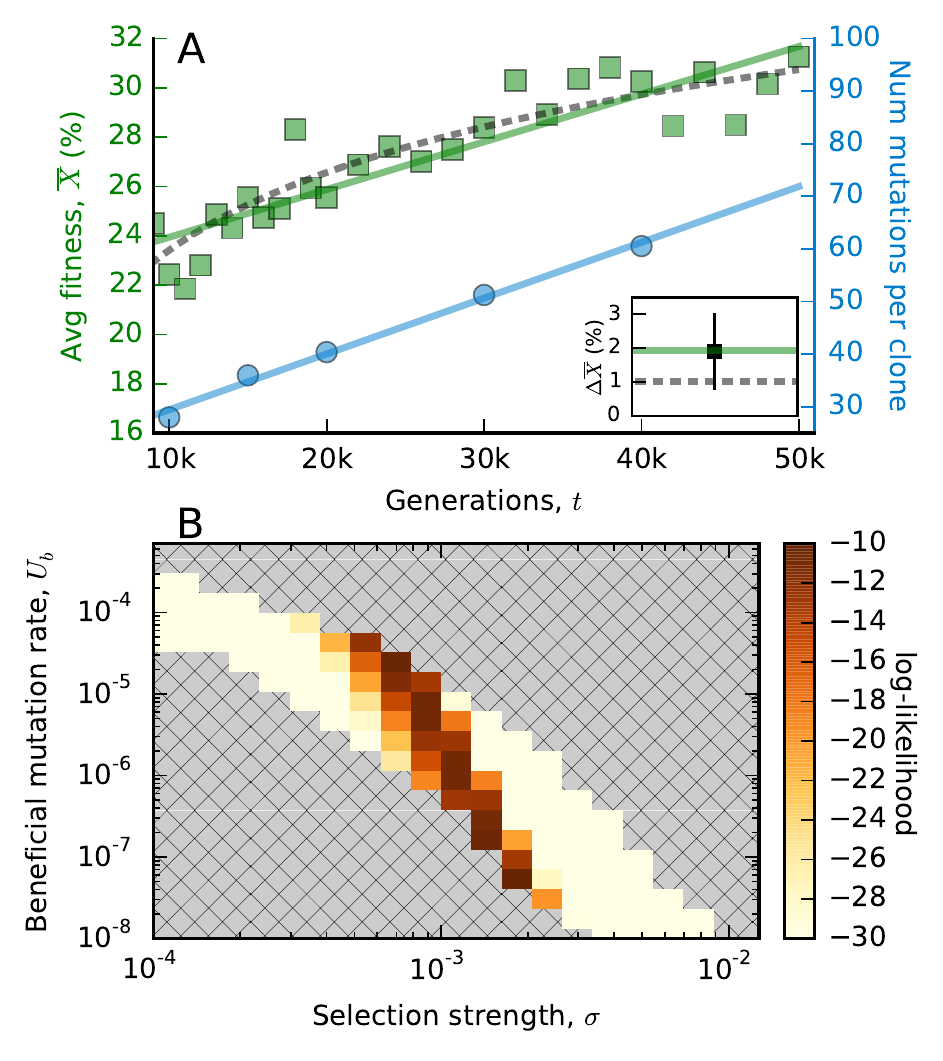}
\caption{Fitting a non-epistatic model to the last 40,000 generations of evolution. (A) The average fitness and mutation trajectories from \fig{fig:trajectory}C, along with the predictions of the non-epistatic curve in \eq{eq:nonepistatic-trajectory} (solid lines). The estimated parameters are $\vo \approx 1.9 \times 10^{-6}$ and $\Ro \approx 1.1 \times 10^{-3}$. For comparison, the best-fit logarithmic trajectory from \eq{eq:wiser-trajectory} is shown by the dashed line. (Inset) Predictions for the average change in fitness between generations 40,000 and 50,000 (lines) compared to the independent measurement from the inset of \fig{fig:trajectory}B (square). (B) The likelihood of the fitness trajectory for a constant, exponential DFE truncated at $s_\mathrm{max} \approx 4 \sigma$, with a $y$-intercept fitted using maximum likelihood. We have only included parameters whose substitution rates are consistent with the observed mutation trajectory ($\Ro-\Utot \leq \partial_t \Mbavg \leq \Ro$). \label{fig:after-10k-fit}}
\end{figure}

Revisiting the original data in \fig{fig:trajectory}, we note that we were initially led to consider models of macroscopic epistasis, rather than a constant DFE, because of the differences between the initial part of the experiment (e.g., before generation 10,000) and the later part of the experiment (e.g., after generation 10,000). The differences in the rate of adaptation are so striking that we can clearly rule out a constant DFE even without a rigorous statistical comparison. In contrast, the deceleration in the fitness trajectory in the final 40,000 generations of evolution is much less pronounced. This leads us to consider a simple statistical question: does the later portion of the experiment actually contain evidence of macroscopic epistasis, or are we simply extrapolating from the strong deceleration in the early part of the fitness trajectory?  

We can frame this statistical question as a model of epistasis with two evolutionary epochs: an initial ``poorly adapted'' phase in the first 10,000 generations followed by a more ``well-adapted'' phase for the remaining 40,000 generations. We do not attempt to model the initial phase, but instead simply assume that the population is subject to some complicated and unspecified model of epistasis that generates the observed data with probability one. This could account for the fact that the large-effect mutations available at the beginning of the experiment might depend on specific details of the ancestral strain or other experimental details. After this initial phase of adaptation is complete, the population enters a second phase of evolution with negligible macroscopic epistasis. In other words, rather than try to fit a single model of a changing DFE to the whole experiment, we neglect the first 10,000 generations and instead try to fit an evolutionary model with a constant DFE to the last 40,000 generations of evolution. 

Assuming a constant DFE implies that the fitness and mutation trajectories after generation 10,000 are given by 
\begin{align} 
\overline{X}(t) = v_0 t + \Xc \, , \quad \overline{M}(t) = R_0 t + M_c \, . \label{eq:nonepistatic-trajectory}
\end{align} 
This fitness trajectory has the same number of nominal parameters as the logarithmic curve in \eq{eq:wiser-trajectory}, although it is important to remember that \eq{eq:wiser-trajectory} carries an implicit functional degree of freedom that was used to obtain the logarithmic trajectory in the first place. \Fig{fig:after-10k-fit} shows that even on a purely curve-fitting level, the non-epistatic fitness trajectory is only marginally less accurate than its epistatic counterpart. We infer an adaptation rate of $v_0 \approx 0.2\%$ per 1000 generations and a substitution rate of $R_0 \approx 1.1$ per 1000 generations. Although this adaptation rate appears to overestimate the fitness gain after generation $40,000$, the more precise fitness assays performed between generations $40,000$ and $50,000$ corroborate the $2\%$ increase (\fig{fig:after-10k-fit}, inset). The fitted values of $\vo$ and $\Ro$ can be used to infer the typical fitness effect of a fixed mutation and a corresponding effective mutation rate based on the relations \begin{align} \seff \approx \frac{v_0}{R_0} \, , \quad v_0 \approx \frac{2 \log (2 N \seff)}{\log^{2} \left( \seff / \Ueff \right)} \, , \end{align} derived in previous theoretical work \citep{desai:fisher:2007}. For the values of $\vo$ and $\Ro$ above, we find a typical fixed fitness effect of order $s_\mathrm{eff} \sim 2 \times 10^{-3}$ and an effective mutation rate of order $U_\mathrm{eff} \sim 2 \times 10^{-6}$.

However, this discussion has so far been based purely on curve-fitting and not on a specific generative model of the dynamics. Using our computational framework, we can evaluate the fit of the two-epoch model more rigorously. To do so, we performed a grid search over combinations of $U$, $\sigma$, and $\Xc$ for a truncated exponential distribution ($s_\mathrm{max} = 4 \sigma$). Recall that there is little power to infer the shape of the DFE in this model; we chose the truncated exponential distribution because its parameters can be directly compared to the best-fit diminishing returns model in \eq{eq:wiser-epistasis}. We find that the best non-epistatic models are statistically consistent with the observed fitness trajectory ($p \approx 0.9$; $\chi^2$ test), and provide only a marginally worse fit than the diminishing returns models above ($\Delta \log \Lambda < -3$). Moreover, this difference in likelihood vanishes completely if we restrict our attention to the last $35,000$ generations of the experiment rather than the last $40,000$. Together, these results suggest that there is limited evidence for macroscopic epistasis in the later portion of the LTEE based on the currently available data. 

\section*{Discussion}

\noindent Genetic reconstructions provide numerous examples of interactions between the fitness effects of individual mutations. The existence of these interactions is hardly surprising, given the physiological and developmental complexity of most organisms. However, the evolutionary implications of these interactions remain controversial. In this study, we used longitudinal data from a long-term evolution experiment in \emph{E. coli} to investigate the evolutionary influence of epistasis in a simple empirical setting. We focused on two basic questions: (1) how do naturally occurring patterns of epistasis alter the rate of fitness increase and the accumulation of new mutations in a constant environment? and (2) what are the simplest models of epistasis that are consistent with the observed data?

The first of these questions is largely descriptive, and has been the focus of previous work in this experimental system and many others \citep{lenski:travisano:1994, wichman:etal:1999, silander:etal:2007, barrick:etal:2009, wiser:etal:2013, kryazhimskiy:etal:2012, kryazhimskiy:etal:2014, perfeito:etal:2014}. The latter question, in contrast, demands a quantitative approach, and must account for the fact that the underlying model of epistasis can only be observed through the filter of population genetic stochasticity and measurement error. In this study, we developed a computational framework to account for these confounding factors, which allows us to quantify the consistency of a predicted fitness trajectory using well-established statistical tools. Combined with analytical results in the weak-mutation limit, we used this framework to investigate the compatibility of several popular models of epistasis.

We found that the shape of a decelerating fitness trajectory on its own provides little power to distinguish between different models of epistasis, including finite-sites models that lack any direct interactions between mutations. This suggests that the underlying ``symmetry group'' or universality class for this observable may be quite large \citep{frank:2014}, which could potentially explain why previous studies have been so successful at fitting the LTEE fitness trajectory with simple epistatic models \citep{sibani:etal:1998, wiser:etal:2013, frank:2014}. However, this symmetry is broken as soon as we include information about the number of mutations that have accumulated, and the combination of fitness and genetic data places much stronger constraints on the set of possible models. Of the simple 2 and 3-parameter models considered here, we found that a variant of the global diminishing returns model proposed by \citet{wiser:etal:2013} provides the best fit to the observed data, although certain systematic errors still remain. These systematic errors, combined with the weak-diminishing returns signal in the first five mutations in the Ara-1 population (File~S1), suggest a degree of caution in interpreting the support for this model.

Instead, we find that the data are equally well explained by a two-epoch model of adaptation, in which an initial burst of macroscopic epistasis is followed by a steady accumulation of mutations under a constant DFE. Although this model offers no insight into the initial (presumably idiosyncratic) phase of adaptation, it provides a more parsimonious explanation for the dynamics in the latter phase of the experiment. Moreover, given that this second phase accounts for three quarters of the present duration of the LTEE and more than half of the accumulated mutations, it could be argued that it provides a better description of the ``typical'' dynamics of adaptation in a constant environment than the initial, epistatic phase of the adaptive walk. Under this hypothesis, the widespread diminishing returns epistasis observed in other experimental systems may simply be a reflection of their comparatively brief duration.

In light of this speculation, it is worth commenting on the population genetic parameters estimated in the second, slower phase of the LTEE. When faced with a constant environment, it is natural to expect that a population will eventually enter a slower phase of adaptation once the most obvious beneficial mutations are exhausted. However, this steady-state is usually assumed to have a negligible beneficial mutation rate and correspondingly simple evolutionary dynamics. In contrast, the scaled mutation rates that we estimate for the LTEE are surprisingly large ($NU \sim 10-100$), and are comparable to rapidly evolving laboratory yeast populations near the beginning of their adaptive walk \citep{frenkel:etal:2014}. Although the fitness effects of these mutations ($s \sim 0.1\%$) fall below the resolution limit of most fitness assays, the effective population size is large enough that the scaled selection strengths are quite large from a population genetic standpoint ($Ns \gtrsim 10^4$). Interestingly, these scaled beneficial mutation rates and selection strengths are sufficiently large that deleterious mutations are expected to have a negligible influence on the rate of adaptation \citep{good:desai:2014}. (We also note that extrapolating these estimates to the mutator lines would suggest that the declining mutation rate observed in Ara-1 \citep{wielgoss:etal:2013} may involve selection for more than just a reduced deleterious load.)

Together, these estimates suggest that even the ``slow'' phase of the LTEE is characterized by rapid adaptation, in which multiple beneficial mutations compete for fixation at the same time. These dynamics are illustrated by the simulated mutation trajectories in \fig{fig:hypothetical-timecourse}, which display an even greater amount of hitchhiking and clonal interference than similar trajectories measured in a recent evolution experiment in yeast \citep{lang:etal:2013}. However, since the individual fitness effects are an order of magnitude smaller, the competition between beneficial mutations occurs over a much longer timescale than is normally observed in experimental evolution. For example, it is not uncommon to find a beneficial mutation that persists for thousands of generations at intermediate frequencies before it accumulates enough additional mutations to sweep to fixation. These transiently stable polymorphisms would suggest adaptive radiation or frequency-dependent selection on a more traditional experimental timescale (e.g. less than $2,000$ generations), but they arise here as a natural consequence of the population genetic process.

Of course, the preceding discussion should be treated with a degree of caution, since our present estimates are based on a limited number of clone sequences and relatively noisy fitness measurements. It is possible that additional data would indicate a departure from the constant adaptation rate in the second phase of the experiment, or reverse some of the systematic errors of the global diminishing returns model. Moreover, even a perfectly resolved fitness and mutation trajectory will likely be consistent with more than one evolutionary model, just as a perfectly resolved fitness trajectory is consistent with multiple mutation trajectories. Our estimates should therefore be viewed as merely consistent with the available data, rather than strongly supported by them. Nevertheless, our results demonstrate that an easy-to-measure genetic observable such as the mutation trajectory can greatly restrict the set of models that are consistent with a measured fitness trajectory. Additional information about the genetic diversity within the population (e.g., measured from pairwise heterozygosity among clones) will likely provide even more power to distinguish between competing hypotheses. The computational framework developed here provides a powerful and flexible method for incorporating this genetic information as it becomes available.

\begin{acknowledgments}
We thank Sergey Kryazhimskiy, Elizabeth Jerison, and Daniel Rice for useful discussions, and Joshua Plotkin, David McCandlish, and two reviewers for helpful comments on the manuscript. This work was supported in part by a National Science Foundation Graduate Research Fellowship, the James S. McDonnell Foundation, the Alfred P. Sloan Foundation, the Harvard Milton Fund, grant PHY 1313638 from the NSF, and grant GM104239 from the NIH. Simulations in this paper were performed on the Odyssey cluster supported by the Research Computing Group at Harvard University. 
\end{acknowledgments}


\begin{thebibliography}{}

\bibitem[\protect\astroncite{Bank et~al.}{2014}]{bank:etal:2014}
{\sc Bank, C., Hietpas, R.~T., Wong, A., Bolon, D.~N., and Jensen, J.~D.} 2014.
\newblock A bayesian mcmc approach to assess the complete distribution of
  fitness effects of new mutations: Uncovering the potential for adaptive walks
  in challenging environments.
\newblock {\em Genetics} 196:841--852.

\bibitem[\protect\astroncite{Barrick et~al.}{2010}]{barrick:etal:2010}
{\sc Barrick, J.~E., Kauth, M.~R., Strelioff, C.~C., and Lenski, R.~E.} 2010.
\newblock \emph{Escherichia coli} rpob mutants have increased evolvability in
  proportion to their fitness defects.
\newblock {\em Mol Biol Evol} 27:1338.

\bibitem[\protect\astroncite{Barrick et~al.}{2009}]{barrick:etal:2009}
{\sc Barrick, J.~E., Yu, D.~S., Yoon, S.~H., Jeong, H., Oh, T.~K., Schneider,
  D., Lenski, R.~E., and Kim, J.~F.} 2009.
\newblock Genome evolution and adaptation in a long-term experiment with
  \emph{Escherichia coli}.
\newblock {\em Nature} 461:1243--1247.

\bibitem[\protect\astroncite{Birky and Walsh}{1988}]{birky:walsh:1988}
{\sc Birky, Jr., C.~W. and Walsh, J.~B.} 1988.
\newblock Effects of linkage on rates of molecular evolution.
\newblock {\em Proc Natl Acad Sci} 85:6414--6418.

\bibitem[\protect\astroncite{Blount et~al.}{2008}]{blount:etal:2008}
{\sc Blount, Z.~D., Borland, C.~Z., and Lenski, R.~E.} 2008.
\newblock Historical contingency and the evolution of a key innovation in an
  experimental population of \emph{Escherichia coli}.
\newblock {\em Proc Natl Acad Sci} 105:7899--7906.

\bibitem[\protect\astroncite{Burch and Chao}{2000}]{burch:chao:2000}
{\sc Burch, C.~L. and Chao, L.} 2000.
\newblock Evolvability of an rna virus is determined by its mutational
  neighbourhood.
\newblock {\em Nature} 406:625--628.

\bibitem[\protect\astroncite{Chou et~al.}{2011}]{chou:etal:2011}
{\sc Chou, H.-H., Chiu, H.-C., Delaney, N.~F., Segr\`{e}, D., and Marx, C.~J.}
  2011.
\newblock Diminishing returns epistasis among beneficial mutations decelerates
  adaptation.
\newblock {\em Science} 332:1190--1192.

\bibitem[\protect\astroncite{Costanzo et~al.}{2010}]{costanzo:etal:2010}
{\sc Costanzo, M., Baryshnikova, A., Bellay, J., Kim, Y., and \emph{et al}}
  2010.
\newblock The genetic landscape of a cell.
\newblock {\em Science} 327:425--431.

\bibitem[\protect\astroncite{de~Visser and Krug}{2014}]{deVisser:krug:2014}
{\sc de~Visser, J. A.~G. and Krug, J.} 2014.
\newblock Empirical fitness landscapes and the predictability of evolution.
\newblock {\em Nature Reviews Genetics} 15:480--490.

\bibitem[\protect\astroncite{Desai and Fisher}{2007}]{desai:fisher:2007}
{\sc Desai, M.~M. and Fisher, D.~S.} 2007.
\newblock Beneficial mutation selection balance and the effect of genetic
  linkage on positive selection.
\newblock {\em Genetics} 176:1759--1798.

\bibitem[\protect\astroncite{Draghi and Plotkin}{2013}]{draghi:plotkin:2013}
{\sc Draghi, J. and Plotkin, J.~B.} 2013.
\newblock Selection biases the prevalence and type of epistasis along adaptive
  trajectories.
\newblock {\em Evolution} 67:3120--3131.

\bibitem[\protect\astroncite{Fisher}{2013}]{fisher:2013}
{\sc Fisher, D.~S.} 2013.
\newblock Asexual evolution waves: fluctuations and universality.
\newblock {\em J Stat Mech} 2013:P01011.

\bibitem[\protect\astroncite{Fisher}{1930}]{fisher:1930}
{\sc Fisher, R.~A.} 1930.
\newblock The distribution of gene ratios for rare mutations.
\newblock {\em Proc Roy Soc Edinburgh} 50:204--219.

\bibitem[\protect\astroncite{Fogle et~al.}{2008}]{fogle:etal:2008}
{\sc Fogle, C.~A., Nagle, J.~L., and Desai, M.~M.} 2008.
\newblock Clonal interference, multiple mutations and adaptation in large
  asexual populations.
\newblock {\em Genetics} 180:2163--2173.

\bibitem[\protect\astroncite{Frank}{2014}]{frank:2014}
{\sc Frank, S.~A.} 2014.
\newblock Generative models versus underlying symmetries to explain biological
  pattern.
\newblock {\em Journal of Evolutionary Biology} 27:1172--1178.

\bibitem[\protect\astroncite{Frenkel et~al.}{2014}]{frenkel:etal:2014}
{\sc Frenkel, E.~M., Good, B.~H., and Desai, M.~M.} 2014.
\newblock The fates of mutant lineages and the distribution of fitness effects
  of beneficial mutations in laboratory budding yeast populations.
\newblock {\em Genetics} 196:1217--1226.

\bibitem[\protect\astroncite{Gerrish and Lenski}{1998}]{gerrish:lenski:1998}
{\sc Gerrish, P. and Lenski, R.} 1998.
\newblock The fate of competing beneficial mutations in an asexual population.
\newblock {\em Genetica} 127:127--144.

\bibitem[\protect\astroncite{Gillespie}{1984}]{gillespie:1984}
{\sc Gillespie, J.} 1984.
\newblock Molecular evolution over the mutational landscape.
\newblock {\em Evolution} 38:1116--1129.

\bibitem[\protect\astroncite{Good and Desai}{2014}]{good:desai:2014}
{\sc Good, B.~H. and Desai, M.~M.} 2014.
\newblock Deleterious passengers in adapting populations.
\newblock {\em Genetics} 198:1183--1208.

\bibitem[\protect\astroncite{Good et~al.}{2012}]{good:etal:2012}
{\sc Good, B.~H., Rouzine, I.~M., Balick, D.~J., Hallatschek, O., and Desai,
  M.~M.} 2012.
\newblock Distribution of fixed beneficial mutations and the rate of adaptation
  in asexual populations.
\newblock {\em Proc. Natl. Acad. Sci.} 109:4950--4955.

\bibitem[\protect\astroncite{Gould}{1989}]{gould:1989}
{\sc Gould, S.~J.} 1989.
\newblock Wonderful life: the Burgess Shale and the nature of history.
\newblock W W Norton, New York.

\bibitem[\protect\astroncite{Haldane}{1927}]{haldane:1927}
{\sc Haldane, J.} 1927.
\newblock The mathematical theory of natural and artificial selection, part v:
  selection and mutation.
\newblock {\em Proc. Camb. Philos. Soc.} 23:828--844.

\bibitem[\protect\astroncite{Hegreness et~al.}{2006}]{hegreness:etal:2006}
{\sc Hegreness, M., Shoresh, N., Hartl, D., and Kishony, R.} 2006.
\newblock An equivalence principle for the incorporation of favorable mutations
  in asexual populations.
\newblock {\em Science} 311:1615--1617.

\bibitem[\protect\astroncite{Jasnos and Korona}{2007}]{jasnos:korona:2007}
{\sc Jasnos, L. and Korona, R.} 2007.
\newblock Epistatic buffering of fitness loss in yeast double deletion strains.
\newblock {\em Nature Genetics} 39:550--554.

\bibitem[\protect\astroncite{Kauffman and Levin}{1987}]{kauffman:levin:1987}
{\sc Kauffman, S. and Levin, S.} 1987.
\newblock Towards a general theory of adaptive walks on rugged landscapes.
\newblock {\em J Theor Biol} 128:11--45.

\bibitem[\protect\astroncite{Kauffman and
  Weinberger}{1989}]{kauffman:weinberger:1989}
{\sc Kauffman, S. and Weinberger, E.~D.} 1989.
\newblock The nk model of rugged fitness landscape and its application to
  maturation of the immune response.
\newblock {\em J Theor Biol} 141:211--245.

\bibitem[\protect\astroncite{Khan et~al.}{2011}]{khan:etal:2011}
{\sc Khan, A.~I., Din, D.~M., Schneider, D., Lenski, R.~E., and Cooper, T.~F.}
  2011.
\newblock Negative epistasis between beneficial mutations in an evolving
  bacterial population.
\newblock {\em Science} 332:1193--1196.

\bibitem[\protect\astroncite{Kingman}{1978}]{kingman:1978}
{\sc Kingman, J. F.~C.} 1978.
\newblock A simple model for the balance between selection and mutation.
\newblock {\em J Appl Prob} 15:1–12.

\bibitem[\protect\astroncite{Kinney et~al.}{2010}]{kinney:etal:2010}
{\sc Kinney, J.~B., Murugana, A., Callan, Jr, C.~G., and Cox, E.~C.} 2010.
\newblock Using deep sequencing to characterize the biophysical mechanism of a
  transcriptional regulatory sequence.
\newblock {\em Proc Natl Acad Sci} 107:9158--9163.

\bibitem[\protect\astroncite{Kryazhimskiy
  et~al.}{2009}]{kryazhimskiy:etal:2009}
{\sc Kryazhimskiy, S., Tkačik, G., and Plotkin, J.~B.} 2009.
\newblock The dynamics of adaptation on correlated fitness landscapes.
\newblock {\em Proc Natl Acad Sci USA} 106:18638--18643.

\bibitem[\protect\astroncite{Kryazhimskiy
  et~al.}{2012}]{kryazhimskiy:etal:2012}
{\sc Kryazhimskiy, S.~K., Rice, D.~P., and Desai, M.~M.} 2012.
\newblock Population subdivision and adaptation in asexual populations of
  saccharomyces cerevisiae.
\newblock {\em Evolution} 66:1931--1941.

\bibitem[\protect\astroncite{Kryazhimskiy
  et~al.}{2014}]{kryazhimskiy:etal:2014}
{\sc Kryazhimskiy, S.~K., Rice, D.~P., Jerison, E.~R., and Desai, M.~M.} 2014.
\newblock Global epistasis makes adaptation predictable despite sequence-level
  stochasticity.
\newblock {\em Science} 344:1519--1522.

\bibitem[\protect\astroncite{Lang and Murray}{2008}]{lang:murray:2008}
{\sc Lang, G.~I. and Murray, A.~W.} 2008.
\newblock Estimating the per-base-pair mutation rate in the yeast
  \emph{Saccharomyces cerevisiae}.
\newblock {\em Genetics} 178:67--82.

\bibitem[\protect\astroncite{Lang et~al.}{2013}]{lang:etal:2013}
{\sc Lang, G.~I., Rice, D.~P., Hickman, M.~J., Sodergren, E., Weinstock, G.~M.,
  Botstein, D., and Desai, M.~M.} 2013.
\newblock Pervasive genetic hitchhiking and clonal interference in forty
  evolving yeast populations.
\newblock {\em Nature} 500:571--574.

\bibitem[\protect\astroncite{Lee et~al.}{2012}]{lee:etal:2012}
{\sc Lee, H., Popodi, E., Tanga, H., and Foster, P.~L.} 2012.
\newblock Rate and molecular spectrum of spontaneous mutations in the bacterium
  \emph{Escherichia coli} as determined by whole-genome sequencing.
\newblock {\em Proc Natl Acad Sci} 109:E2774--E2783.

\bibitem[\protect\astroncite{Lenski et~al.}{1991}]{lenski:etal:1991}
{\sc Lenski, R.~E., Rose, M.~R., Simpson, S.~C., and Tadler, S.~C.} 1991.
\newblock Long-term experimental evolution in escherichia coli. i. adaptation
  and divergence during 2,000 generations.
\newblock {\em The American Naturalist} 138:1315--1341.

\bibitem[\protect\astroncite{Lenski and
  Travisano}{1994}]{lenski:travisano:1994}
{\sc Lenski, R.~E. and Travisano, M.} 1994.
\newblock Dynamics of adaptation and diversification: A 10,000-generation
  experiment with bacterial populations.
\newblock {\em Proc Natl Acad Sci USA} 91:6808--6814.

\bibitem[\protect\astroncite{MacLean et~al.}{2010}]{maclean:etal:2010}
{\sc MacLean, R.~C., Perron, G.~G., and Gardner, A.} 2010.
\newblock Diminishing returns from beneficial mutations and pervasive epistasis
  shape the fitness landscape for rifampicin resistance in \emph{Pseudomonas
  aeruginosa}.
\newblock {\em Genetics} 186:1345--1354.

\bibitem[\protect\astroncite{Matic et~al.}{1997}]{matic:etal:1997}
{\sc Matic, I., Radman, M., Taddei, F., Picard, B., Doit, C., Bingen, E.,
  Denamur, E., and Elion, J.} 1997.
\newblock Highly variable mutation rates in commensal and pathogenic
  \emph{Escherichia coli}.
\newblock {\em Science} 277:1833--1834.

\bibitem[\protect\astroncite{McCandlish
  et~al.}{2014a}]{mccandlish:epstein:plotkin:2014}
{\sc McCandlish, D., Epstein, C., and Plotkin, J.~B.} 2014a.
\newblock The inevitability of unconditionally deleterious substitutions during
  adaptation. evolution.
\newblock  68:1351--1365.

\bibitem[\protect\astroncite{McCandlish
  et~al.}{2014b}]{mccandlish:otwinowski:plotkin:2014}
{\sc McCandlish, D.~M., Otwinowski, J., and Plotkin, J.~B.} 2014b.
\newblock On the role of epistasis in adaptation.
\newblock {\em arXiv} .

\bibitem[\protect\astroncite{Miller et~al.}{2011}]{miller:etal:2011}
{\sc Miller, C.~R., Joyce, P., and Wichman, H.~A.} 2011.
\newblock Mutational effects and population dynamics during viral adaptation
  challenge current models.
\newblock {\em Genetics} 187:185--202.

\bibitem[\protect\astroncite{Nagel et~al.}{2012}]{nagel:etal:2012}
{\sc Nagel, A.~C., Joyce, P., Wichman, H.~A., and Miller, C.~R.} 2012.
\newblock Stickbreaking: A novel fitness landscape model that harbors epistasis
  and is consistent with commonly observed patterns of adaptive evolution.
\newblock {\em Genetics} 190:655--667.

\bibitem[\protect\astroncite{Orr}{2002}]{orr:2002}
{\sc Orr, H.} 2002.
\newblock The population genetics of adaptation: the adaptation of dna
  sequences.
\newblock {\em Evolution} 56:1317--1330.

\bibitem[\protect\astroncite{Perfeito et~al.}{2014}]{perfeito:etal:2014}
{\sc Perfeito, L., Sousa, A., Bataillon, T., and Gordo, I.} 2014.
\newblock Rates of fitness decline and rebound suggest pervasive epistasis.
\newblock {\em Evolution} 68:150--162.

\bibitem[\protect\astroncite{Quandt et~al.}{2014}]{quandt:etal:2014}
{\sc Quandt, E.~M., deatherage, D.~E., Ellington, A.~D., Georgiou, G., and
  Barrick, J.~E.} 2014.
\newblock Recursive genomewide recombination and sequencing reveals a key
  refinement step in the evolution of a metabolic innovation in
  \emph{Escherichia coli}.
\newblock {\em Proc Natl Acad Sci} 111:2217--2222.

\bibitem[\protect\astroncite{Rozen and Lenski}{2000}]{rozen:lenski:2000}
{\sc Rozen, D.~E. and Lenski, R.~E.} 2000.
\newblock Long-term experimental evolution in escherichia coli. viii. dynamics
  of a balanced polymorphism.
\newblock {\em American Naturalist} 155:24--35.

\bibitem[\protect\astroncite{Schiffels et~al.}{2011}]{schiffels:etal:2011}
{\sc Schiffels, S., Sz\"{o}ll\"{o}si, G., Mustonen, V., and L\"{a}ssig, M.}
  2011.
\newblock Emergent neutrality in adaptive asexual evolution.
\newblock {\em Genetics} 189:1361--1375.

\bibitem[\protect\astroncite{Segr\`{e} et~al.}{2005}]{segre:etal:2005}
{\sc Segr\`{e}, D., Deluna, A., Church, G., and Kishony, R.} 2005.
\newblock Modular epistasis in yeast metabolism.
\newblock {\em Nat Genet} 37:77--83.

\bibitem[\protect\astroncite{Sibani et~al.}{1998}]{sibani:etal:1998}
{\sc Sibani, P., Brandt, M., and Alstr{\o}m, P.} 1998.
\newblock Evolution and extinction dynamics in rugged fitness landscapes.
\newblock {\em Intl J Mod Phys} 12:361--391.

\bibitem[\protect\astroncite{Silander et~al.}{2007}]{silander:etal:2007}
{\sc Silander, O.~K., Tenaillon, O., and Chao, L.} 2007.
\newblock Understanding the evolutionary fate of finite populations: The
  dynamics of mutational effects.
\newblock {\em PLoS Biol} 5:e94.

\bibitem[\protect\astroncite{St~Onge et~al.}{2007}]{st-onge:etal:2007}
{\sc St~Onge, R.~P., Mani, R., Oh, J., Proctor, M., and \emph{et al}} 2007.
\newblock Systematic pathway analysis using high-resolution fitness profiling
  of combinatorial gene deletions.
\newblock {\em Nature Genetics} 39:199 -- 206.

\bibitem[\protect\astroncite{Tenaillon et~al.}{2012}]{tenaillon:etal:2012}
{\sc Tenaillon, O., Rodríguez-Verdugo, A., Gaut, R.~L., McDonald, P., Bennett,
  A.~F., Long, A.~D., and Gaut, B.~S.} 2012.
\newblock The molecular diversity of adaptive convergence.
\newblock {\em Science} 335:457--461.

\bibitem[\protect\astroncite{Weinreich et~al.}{2006}]{weinreich:etal:2006}
{\sc Weinreich, D.~M., Delaney, N.~F., DePristo, M.~A., and Hartl, D.~L.} 2006.
\newblock Darwinian evolution can follow only very few mutational paths to
  fitter proteins.
\newblock {\em Science} 312:111--114.

\bibitem[\protect\astroncite{Wichman et~al.}{1999}]{wichman:etal:1999}
{\sc Wichman, H.~A., Badgett, M.~R., Scott, L.~A., Boulianne, C.~M., and Bull,
  J.~J.} 1999.
\newblock Different trajectories of parallel evolution during viral adaptation.
\newblock {\em Science} 285:422--424.

\bibitem[\protect\astroncite{Wielgoss et~al.}{2011}]{wielgoss:etal:2011}
{\sc Wielgoss, S., Barrick, J.~E., Tenaillon, O., Cruveiller, S.,
  Chane-Woon-Ming, B., M\'{e}digue, C., Lenski, R.~E., Schneider, D., and
  Andrews, B.~J.} 2011.
\newblock Mutation rate inferred from synonymous substitutions in a long-term
  evolution experiment with escherichia coli.
\newblock {\em G3 (Bethesda)} 1:183--186.

\bibitem[\protect\astroncite{Wielgoss et~al.}{2013}]{wielgoss:etal:2013}
{\sc Wielgoss, S., Barrick, J.~E., Tenaillon, O., Wiser, M.~J., Dittmar, J.,
  Cruveiller, S., Chane-Woon-Ming, B., M\'{e}digue, C., Lenski, R.~E., and
  Schneider, D.} 2013.
\newblock Mutation rate dynamics in a bacterial population reflect tension
  between adaptation and genetic load.
\newblock {\em Proc Natl Acad Sci} 110:222–227.

\bibitem[\protect\astroncite{Wiser et~al.}{2013}]{wiser:etal:2013}
{\sc Wiser, M.~J., Ribeck, N., and Lenski, R.~E.} 2013.
\newblock Long-term dynamics of adaptation in asexual populations.
\newblock {\em Science} 342:1364--1367.

\bibitem[\protect\astroncite{Woods et~al.}{2011}]{woods:etal:2011}
{\sc Woods, R.~J., Barrick, J.~E., Cooper, T.~F., Shrestha, U., Kauth, M.~R.,
  and Lenski, R.~E.} 2011.
\newblock Second-order selection for evolvability in a large \emph{Escherichia
  coli} population.
\newblock {\em Science} 331:1433--1436.

\bibitem[\protect\astroncite{Wright}{1932}]{wright:1932}
{\sc Wright, S.} 1932.
\newblock The roles of mutation, inbreeding, crossbreeding and selection in
  evolution.
\newblock {\em In} Proceedings of the VI International Congress of Genetics,
  pp. 356--366.

\bibitem[\protect\astroncite{Zhu et~al.}{2014}]{zhu:etal:2014}
{\sc Zhu, Y., Siegal, M.~L., Hall, D.~W., and Petrov, D.~A.} 2014.
\newblock Precise estimates of mutation rate and spectrum in yeast.
\newblock {\em Proc Natl Acad Sci} 111:E2310--E2318.

\end{thebibliography}

\section*{Appendix}

\paragraph*{Fitness trajectories.}

The fitnesses of the LTEE strains were calculated from \eq{eq:fitness} using the raw competition assays reported by \citet{wiser:etal:2013}. The quantity $f/(1-f)$ was estimated from the ratio of red and white colonies when plated on arabinose media, and the duration each competition was $\Delta t = \log_2 (100) \approx 6.6$ generations. The fitness gains between generation $40,000$ and $50,000$ were calculated in a similar manner, but with a longer competition time of $\Delta t = 3 \log_2(100) \approx 19.9$ generations. Subsequent analysis of the fitness trajectory was restricted to the six non-mutator populations with complete fitness measurements: Ara$-5$, Ara$-6$, Ara$+1$, Ara$+2$, Ara$+4$, and Ara$+5$ \citep{wiser:etal:2013}. These populations were chosen because they have complete fitness trajectories that can be reasonably expected to evolve under the same population genetic model. Notably, this subset excludes both the citrate-metabolizing population (Ara$-3$) studied by \citet{blount:etal:2008}, as well as the crossfeeding population (Ara$-2$) studied by \citet{rozen:lenski:2000}. 

\paragraph*{Fitness effects of individual mutations.} The fitness effects of the five mutations in \fig{fig:khan-trajectory} were calculated from the raw competition assays reported by \citet{khan:etal:2011}. The fitness effect $s$ was defined as the difference between the fitness of the mutant and background genotypes, which were estimated from the competition assays using the same procedure as above.

\paragraph*{The mutation trajectory.} The total number of genetic changes was estimated from the DNA sequences of clones analyzed by \citet{barrick:etal:2009} and \citet{wielgoss:etal:2011}. The finalized mutation calls for the clones in \citet{barrick:etal:2009} were obtained from supplementary tables 1 and 2 of that work, and the mutation calls for the clones in  \citet{wielgoss:etal:2011} were obtained from the supplementary data files available at \url{http://barricklab.org/twiki/pub/Lab/SupplementLongTermMutationRates/long-term_mutation_rates.zip}. To obtain a more densely sampled mutation trajectory, we included clones from populations that were excluded from the fitness trajectory analysis above. This includes seven clones sampled from the Ara$-1$ population prior to the spread of the mutator phenotype, and two clones sampled from the Ara$-3$ population prior to the spread of the citrate-metabolizing phenotype. The complete list of included clones is given in \tbl{table:clone-list}.

\paragraph*{Population genetic simulations.} Simulated fitness and mutation trajectories were obtained from a forward-time algorithm designed to mimic the serial transfer protocol of the LTEE. Between each transfer, lineages are assumed to expand clonally for $\log_2(100) \approx 6.64$ generations at a deterministic exponential growth rate $r = r_0 + X$, where $X$ is the fitness relative to the ancestor. At the transfer step, the population is diluted 100-fold (with Poisson sampling noise) to $N_b = 5 \times 10^6$ individuals. Mutations accumulate at a constant rate $U$ during the growth phase, but we assumed that they do not significantly influence the fitness of the individual until the next transfer cycle. Thus, mutation was approximated by assuming that each individual has a probability $6.64 \cdot U$ of gaining a mutation at the end of a transfer step, with additive fitness effects drawn from the genotype-specific DFE, $\rho(s|\vec{g})$. For the finite-sites model, a discrete ancestral DFE was initialized by drawing $L$ fitness effects from the continuous distribution $\rhoo(s)$, with the same realization shared across replicate lines. A copy of our implementation in \verb|C++| is available at \verb|https://github.com/benjaminhgood/ltee_inference|. 

\paragraph*{Likelihood estimation.} The likelihood of each parameter combination was estimated from simulations using \eq{eq:approx-likelihood}. To speed computation, we simulated 18 replicate populations and generated $n = 10,000$ different 6-population averages by bootstrap resampling. The scaled likelihood $\Lambda$, which differs from $p(\Xavg_\mathrm{obs},\dxavg_\mathrm{obs}|\theta)$ by a constant factor, was defined as
\begin{align}
\begin{aligned}
\Lambda \equiv \sum_{i=1}^n & \exp \left[ - \frac{\sum_{\{t_k\}} \left( \overline{X}_\mathrm{obs}(t_k)-\overline{X}_i(t_k) \right)^2}{2 \sigmax^2} \right] \\
	& \times \exp \left[ - \frac{(\dxavg_\mathrm{obs} - \dxavg_i)^2}{2 \sigmadx^2} \right] \, ,
\end{aligned}
\end{align}
where the measurement uncertainties $\sigmax \approx 1.4\%$ and $\sigmadx \approx 0.4\%$ were estimated from \fig{fig:trajectory}.

\paragraph*{Statistical tests.} The consistency of each parameter combination was assessed using a $\chi^2$ goodness-of-fit test. We simulated 18 replicate populations to estimate $\Xavg(t)$ and $\dxavg$, and we generated $n=10,000$ different 6-population averages by bootstrap resampling these replicates and adding unbiased Gaussian measurement noise with $\sigmax = 1.4\%$ and $\sigmadx = 0.4\%$. The $p$-value is then approximated by
\begin{align}
\mathrm{Pr}\left[ \chi^2 > \chi^2_\mathrm{obs} | \theta \right] \approx \frac{1}{n} \sum_{i=1}^n \theta \left( \chi_i^2 - \chi_\mathrm{obs}^2 \right)
\end{align}
where $\theta(x)$ is the Heaviside step function and $\chi_i^2$ is the mean squared error,
\begin{align}
\chi_i^2 \equiv \frac{\sum_{t_k} \left( \Xavg_i(t_k) - \Xavg(t_k) \right)^2}{\sigmax^2} + \frac{(\dxavg_i - \dxavg)^2}{\sigmadx^2} \, .
\end{align}
The posterior predictive $p$-value for the entire model is then defined by
\begin{align}
p = \frac{\int \mathrm{Pr}[\chi > \chi_\mathrm{obs} | \theta] p(\overline{X}_\mathrm{obs}|\theta) p(\theta) \, d\theta}{\int p(\overline{X}_\mathrm{obs}|\theta) p(\theta)} \, d\theta \, ,
\end{align}
where $p(\theta)$ is the prior distribution of parameter values.

\onecolumngrid

\setcounter{figure}{0} 
\makeatletter 
\renewcommand{\thefigure}{S\@arabic\c@figure} 
\makeatother

\setcounter{table}{0} 
\makeatletter 
\renewcommand{\thetable}{S\@arabic\c@table} 
\makeatother

\makeatletter 
\def\tagform@#1{\maketag@@@{(S\arabic{section}.\ignorespaces#1\unskip\@@italiccorr)}}
\makeatother
\makeatletter 

\newcommand{\seq}[1]{Eq.~(S\arabic{section}.\ref{#1})}
\newcommand{\seqs}[2]{Eqs.~(S\arabic{section}.\ref{#1}) and (S\arabic{section}.\ref{#2})}

\clearpage
 
\newpage
\section*{Supplemental Information}

\section{The neutral mutation rate}
\label{appendix:neutral-mutation-rate}
\setcounter{equation}{0}

\noindent To predict the patterns of fitness and mutation accumulation in the LTEE, our population genetic model utilizes the key approximation,
\begin{align}
U \rho(s) \approx \Un \delta(s) + \Ub \rhob(s) \, ,
\end{align}
which partitions the DFE into a set of strongly beneficial ``driver'' mutations and a collection of nearly neutral ``passengers'' \citep{schiffels:etal:2011, good:desai:2014}. By assumption, the driver mutations set the important evolutionary timescales in the system (e.g., the rate of adaptation and the coalescence timescale), while the passenger mutations constitute a perturbative correction. In recent theoretical work, we have shown that this approximation is accurate when the mutation rate is small compared to the relevant fitness differences in the population \citep{good:desai:2014}, which is expected to be the case for the non-mutator lines in the LTEE. Note that as defined above, the passenger portion of the DFE is comprised not only of truly neutral mutations ($|s| \lesssim 1/N$), but also those mutations that approach the neutral substitution rate by hitchhiking with the beneficial drivers. As such, $\Un$ has an implicit dependence on both $N$ and $\Ub \rhob(s)$. 

Since $\Un$ is an effective parameter, it cannot be measured directly. Instead, it must be self-consistently inferred from the data along with the other model parameters. In principle, this is straightforward: the passenger mutations do not influence the fitness trajectory by construction, so the neutral mutation rate can be estimated from a regression of the mutation trajectory residuals, $\Mavg_\mathrm{obs}(t) - \Mbavg(t)$. Nevertheless, we will find it useful to place some crude bounds on $\Un$, both to limit the range of the regression and to allow for back-of-the-envelope arguments that do not require the full precision of our computational inference scheme. To establish these bounds, we will make use of the fact that the neutral mutation rate cannot exceed the total per genome mutation rate or the observed mutation accumulation rate. 
    
Unlike the $\Un$, the total genomic mutation rate is a directly measurable quantity, although these measurements can be confounded by dependence on the genetic background \citep{matic:etal:1997}, genomic heterogeneity \citep{lang:murray:2008}, and the role of complex mutational events. In previous work, \citet{wielgoss:etal:2011} estimated that the total point mutation rate in the LTEE strain is approximately $U_\mathrm{point} \sim 4 \times 10^{-4}$, based on the number of synonymous mutations that have accumulated over the course of the experiment. A mutation accumulation study in a different strain of \emph{E. coli} found that small indels occur approximately an order of magnitude less freuqently than point mutations \citep{lee:etal:2012}, or $U_\mathrm{indel} \approx 0.1 U_\mathrm{point}$. This is comparable to observations in other organisms such as yeast \citep{zhu:etal:2014}. In comparison, much less is known about the mutation rate for larger indels and other chromosomal rearrangements, such as those arising from insertion sequence (IS) elements. This uncertainty is particularly problematic for the LTEE, since these complex mutations constitute a substantial fraction of the observed mutation trajectory. A recent mutation accumulation study in a different strain of \emph{E. coli} estimated that the total rate of IS events in their background is approximately $U_{IS} \sim 3 \times 10^{-4}$. In the LTEE lines, $\sim 100$ rearrangements were observed across all twelve populations by generation 40,000, which suggests that the mutation rate to neutral IS mutations is bounded by $U_{IS,n} \leq 2 \times 10^{-4}$. However, this is likely to be an overestimate, since some of these mutations are probably beneficial. For example, in the single population where timecourse information is available (Ara-1), the accumulation of rearrangements slows significantly after 10,000 generations. If we restrict our attention to the mutations that have accumulated after generation 10,000, the bound on $U_{IS,n}$ drops to $6 \times 10^{-5}$, which is more in line with the rate of small indels. Note, however, that all of these estimates are based on extremely limited data, and are highly susceptible to statistical fluctuations and other ascertainment biases. In the absence of more precise estimates, we decided to employ the combined bound,
\begin{align}
\Un \leq U_\mathrm{point} + U_\mathrm{indel} + U_\mathrm{IS,n} \lesssim 7 \times 10^{-4} \, ,
\end{align}
with the hope that any underestimation of the rate of complex mutational events is balanced by an overestimation of the neutral fraction of point mutations. 

\section{The ``running out of mutations'' model}
\label{appendix:running-out-of-mutations}
\setcounter{equation}{0}

\noindent In the main text, we analyzed the patterns of fitness and mutation accumulation in a simple ``running out of mutations'' model of macroscopic epistasis. Here, we present our model in more detail, and show how the continuum analysis in the main text emerges from a model which is fundamentally based on a finite number of sites. In the most general form of this model, we consider a collection of $\Lb$ sites with fitness effects $\{ s_i \}_{i=1}^{\Lb}$ and target sizes $\{ \mu_i \}_{i=1}^{\Lb}$, from which we can define a \emph{joint} distribution of target sizes and fitness effects,
\begin{align}
f_0(\mu,s) = \frac{1}{\Lb} \sum_{i=1}^{\Lb} \delta(\mu-\mu_i) \delta(s-s_i) \, .
\end{align}
Here, $f_0(\mu,s)$ can be interpreted as the probability density that a randomly drawn site has a target size $\mu$ and fitness effect $s$. Similarly, the marginal distribution $f_0(s) = \int f_0(\mu,s) \, d\mu$ can be interpreted as the probability density that a randomly drawn site has fitness effect $s$, independent of the target size. Note that $f_0(s)$ differs from the traditional DFE, $\rhoo(s)$, which is the probability density that a randomly drawn \emph{mutation} has fitness effect $s$. Since mutations will be biased towards sites with larger target sizes, the DFE corresponds to the weighted integral, 
\begin{align}
\Ub \rhoo(s) = \Lb \int \mu f_0(\mu,s) \, d\mu \ . 
\end{align}
So far, our discussion has been purely notational. The content of this model comes from the assumption that, once a given site has mutated, further mutations at that site are no longer beneficial. For example, after a loss-of-function mutation in a particular pathway, further loss-of-function mutations in the same pathway are expected to have little to no effect ($s \approx 0$), while the reversion will restore the original function of the gene ($s < 0$). In the large population limit, these mutated sites will behave as if they are effectively removed from the beneficial portion of the DFE. Based on this intuition, we define a collection of indicator variables $\{ I_i(t) \}_{i=1}^{\Lb}$, where $I_i(t) = 1$ if a mutation at site $i$ has fixed by time $t$. We can use these indicator variables to define a time-dependent version of $f_0(\mu,s)$,
\begin{align}
f(\mu,s,t) = \frac{1}{\Lb} \sum_{i=1}^{\Lb} [1-I_i(t)] \delta(\mu-\mu_i) \delta(s-s_i) \, ,
\end{align}
which satisfies the initial condition $f(\mu,s,0) = f_0(\mu,s)$. In the SSWM limit, $\pfix(s) \approx 2s$ is independent of $f(\mu,s,t)$, so that the latter evolves as
\begin{align}
\partial_t \langle f(\mu,s,t) \rangle = - 2 N \mu s \langle f(\mu,s,t) \rangle \, , \label{eq:modular-epistasis-dfe-equation}
\end{align}
or
\begin{align}
\langle f(\mu,s,t) \rangle = f_0(\mu,s) e^{-2 N \mu s t} \, , \label{eq:modular-epistasis-dfe-solution} 
\end{align} 
where $\langle f(\mu,s,t) \rangle$ denotes the expectation value over $\{ I_i(t) \}_{i=1}^{\Lb}$. It is similarly straightforward to show that the average fitness and mutation trajectories, $\Xavg(t) = \sum s_i \langle I_i(t) \rangle$ and $\Mavg(t) = \sum \langle I_i(t) \rangle$, evolve as
\begin{subequations}
\label{eq:modular-epistasis-trajectories}
\begin{align}
\partial_t \Xavg(t) & = \int 2 N \Lb \mu s^2 e^{-2 N \mu s t} f_0(\mu,s) \, d\mu \, ds \, , \\
\partial_t \Mbavg(t) & = \int 2 N \Lb \mu s e^{-2 N \mu s t} f_0(\mu,s) \, d\mu \, ds \, .
\end{align}
\end{subequations}
Thus, our model corresponds to the strong-selection ($Ns \to \infty$) limit of the weak-mutation model analyzed by \citet{mccandlish:otwinowski:plotkin:2014}. We recover \threeeqs{eq:dfe-evolution}{eq:trajectory-equations}{eq:sswm-dfe-evolution} in the main text by demanding that all sites have the same target size $\mu = \Ub/\Lb$, so that 
\begin{align}
f_0(\mu,s) = \delta\left( \mu - \frac{\Ub}{\Lb} \right) \rhoo(s) \, .
\end{align}
Note that in the present framework, our pseudo-continuous DFEs exist purely for notational convenience, since the set of available mutations is always bounded in practice. However, in certain cases it will be computationally convenient to consider the large $\Lb$ limit of these equations, in which sums over discrete numbers of sites are replaced with integrals over a continuous distribution. We have constructed our notation in such a way that this limit requires no modification of our equations, with the implicit caveat that we can only consider distributions for which this limit is well-defined. This excludes pathological cases like heavy-tailed DFEs, which lead to singular dynamics in the $\Lb \gg 1$ limit.  

\subsection*{Derivation of \eq{eq:wiser-dfe}}

\noindent In this section, we show how one can obtain a logarithmic fitness trajectory from a finite sites model with an ancestral DFE, 
\begin{align}
\rhoo(s) \propto \begin{cases}
s^{-2} e^{-s/\sigma} & \text{if $s > \epsilon \sigma$,} \\
0 & \text{else,} 
\end{cases}
\label{eq:appendix-finite-weird-dfe}
\end{align}
where $\epsilon$ is a small parameter. We will make use of the asymptotic expansion
\begin{align}
I_p(\epsilon) \equiv \int_\epsilon^\infty \xi^{-p} e^{-\xi} \, d\xi \sim \begin{cases}
\frac{\epsilon^{1-p}}{p-1} & \text{if $p > 1$,} \\
\log \left( \frac{1}{\epsilon} \right) & \text{if $p=1$,} \\
\Gamma(1-p) & \text{if $p < 1$.} 
\end{cases}
\end{align}
The normalizing constant for the DFE is therefore given by
\begin{align}
C = \frac{\sigma}{ \int_\epsilon^\infty \xi^{-2} e^{-\xi} \, d\xi} \approx \epsilon \sigma \, , 
\end{align}
and the mean and mean-squared fitness effects are 
\begin{align}
\langle s \rangle & = \epsilon \sigma \int_{\epsilon}^\infty \xi^{-1} e^{-\xi} \, d\xi \approx \epsilon \sigma \log \left( \frac{1}{\epsilon} \right) \, ,  \\ 
\langle s^2 \rangle & = \epsilon \sigma^2 \int_{\epsilon}^\infty e^{-\xi} \, d\xi \approx \epsilon \sigma^2 \, .
\end{align}
Substituting \eq{eq:sswm-dfe-evolution} into \eq{eq:trajectory-equations}, we find that the fitness trajectory is given by 
\begin{align}
\partial_t \Xavg(t) & = \int 2 N \Ub s^2 e^{-2 N U s t / L} \rhoo(s) = 2 N \Ub \epsilon \sigma^2 \int_{\epsilon}^\infty \exp \left[ - \left( \frac{2 N \Ub \sigma t}{L} + 1 \right) \xi \right] \, d\xi \, , \nonumber \\
	& \approx \left( \frac{2 N \Ub \epsilon \sigma^2}{1+\frac{2 N \Ub \sigma t}{L}} \right)  e^{- \frac{2N\Ub \sigma \epsilon t}{L} } \, ,
\end{align}
which yields a logarithmic fitness trajectory
\begin{align}
\Xavg(t) = \left( L \sigma \epsilon \right) \log \left( 1 + \frac{2 N \Ub \epsilon \sigma^2 t}{L \sigma \epsilon} \right) \, ,
\end{align}
provided that $t \ll L/(2N\Ub \sigma \epsilon)$. When $t \approx L/(2N\Ub\sigma \epsilon)$, we have $\Xavg(t) \approx \Xavg(\infty)$, so we can also write this condition in the form $\Xavg(t) \ll \Xavg(\infty)$. Even after fitting $\Xc = L \sigma \epsilon$ and $\vo = 2 N \Ub \epsilon \sigma^2$, there is still sufficient freedom to choose the parameters so that this condition holds for any finite time $t_\mathrm{max}$ or fitness $\Xavg(t_\mathrm{max})$. 


\subsection*{Heterogeneous target sizes}

\noindent It is also useful to investigate the consequences of the running out of mutations model when we relax the uniform target size assumption. This becomes unwieldy in our original DFE notation, but simplifies considerably if we change variables from the target size $\mu$ to the substitution rate $R = 2 N \mu s$. From the change of variables theorem, this induces induces a related joint distribution,
\begin{align}
g_0(R,s) = \frac{f_0(R/2Ns, s)}{2Ns} \, ,
\end{align}
which allows us to rewrite the fitness and mutation trajectories in the form 
\begin{subequations}
\label{eq:transformed-modular-epistasis-trajectories}
\begin{align}
\partial_t \Xavg(t)  & = \Lb \int s R e^{-R t} g_0(R,s) \, dR \, ds = \Lb \int h(R) R e^{-R t} g_0(R) \, dR \, , \\
\partial_t  \Mbavg(t) & = \Lb \int R e^{-R t} g_0(R,s) \, dR \, ds = \Lb \int R e^{-R t} g_0(R) \, dR \, .
\end{align}
\end{subequations}
Here, $g_0(R) = \int g_0(s,R) \, ds$ is the marginal distribution of $R$ and $h(R) = g_0(R)^{-1} \int s g_0(R,s) \, ds$ is the conditional mean of $s$ given $R$. Thus, it is easy to see that for a \emph{fixed} $h(R)$, the relationship between $\Xavg(t)$ and $\Mbavg(t)$ is completely determined. For example, we recover \eq{eq:finite-mutation-trajectory} in the main text when $h(R) \propto R$, even if the distribution of target sizes is not completely uniform. However, it is also clear that if we are allowed to tune $h(R)$ and $g_0(R)$ independently, then it is possible to fit both $\Xavg(t)$ and $\Mavg(t)$ simultaneously with the inverse Laplace transforms
\begin{subequations}
\begin{align}
g_0(R) & \propto \frac{\mathcal{L}^{-1} \left\{ \partial_t \Mbavg \right\}}{\Lb R} \, , \\
h(R) & \propto \frac{\mathcal{L}^{-1} \{ \partial_t \Xavg \}}{\mathcal{L}^{-1} \{ \partial_t \Mbavg \}} \, ,  
\end{align}
\end{subequations}
subject to the same technical restrictions on $\Xavg(t)$ and $\Mbavg(t)$ that we encountered in the text. See also related results by \citep{mccandlish:otwinowski:plotkin:2014}, who study similar questions while relaxing the strong selection requirement. 

For example, we can reproduce both the fitness and mutation trajectories of the global diminishing returns model in \eq{eq:wiser-epistasis} by choosing
\begin{align}
g_0(R) & \propto \begin{cases}
R^{-3/2} e^{-R/\tilde{R}} & \text{if $R > \epsilon \tilde{R}$} \, , \\
0 & \text{else,}
\end{cases} \\
h(R) & = \tilde{s} \left( \frac{R}{\tilde{R}} \right)^{1/2} \, , 
\end{align}
where 
\begin{align}
\tilde{R} = \frac{2N\Ub \langle s \rangle \sfixed}{\Xc} \, , \quad c = \sfixed \, , \quad \Lb = \frac{2 \Xc}{\sfixed \sqrt{\epsilon}} \, ,
\end{align}
and $\epsilon \ll 1$ is a small parameter chosen to maintain normalization. We can achieve this by choosing a joint distribution for $R$ and $s$ of the form
\begin{align}
g_0(R,s) \propto \begin{cases}
\delta\left[ s - c \left( \frac{R}{\tilde{R}} \right)^{1/2} \right] R^{-3/2} e^{-R/\tilde{R}} & \text{if $R > \epsilon \tilde{R}$,} \\
0 & \text{else.}
\end{cases}
\end{align}
Switching back to $\mu$ and $s$, we have
\begin{align}
f_0(\mu,s) & \propto \begin{cases} (2Ns) \cdot \delta \left[ s - c \left( \frac{2Ns \mu}{\tilde{R}} \right)^{1/2} \right] (2Ns\mu)^{-3/2} e^{-2Ns \mu / \tilde{R}} & \text{if $2Ns\mu > \epsilon \tilde{R}$,} \\
0 & \text{else,}
\end{cases} \nonumber \\
    & \propto \begin{cases}
     (2Ns)^{-1/2} \cdot \delta \left[ \mu - \frac{\tilde{R} s}{2Nc^2} \right] \mu^{-3/2} e^{-2 N s \mu / \tilde{R}} & \text{if $2Ns\mu > \epsilon \tilde{R}$,} \\
     0 & \text{else,}
\end{cases} \nonumber \\
	& \propto \begin{cases}
	 \delta \left[ \mu - \frac{\Ub \langle s \rangle s}{\Xc c} \right] s^{-2} \exp \left[ - \left( \frac{s}{c} \right)^{2} \right] & \text{if $s > c\sqrt{\epsilon}$,} \\
	 0 & \text{else,} 
\end{cases}
\end{align}
where the normalization factor is
\begin{align}
C = \frac{1}{\int_{c \sqrt{\epsilon}}^\infty s^{-2} e^{-(s/c)^2} \, ds} = \frac{2c}{\int_{\epsilon}^\infty \xi^{-3/2} e^{-\xi} \, d\xi} = c \epsilon^{1/2} \, .
\end{align}
In terms of the traditional DFE, $\rhoo(s)$, we have
\begin{align}
\rhoo(s) \propto \int \mu f_0(\mu,s) \, d\mu \propto \begin{cases}
 s^{-1} e^{-(s/c)^2}  & \text{if $s > c \sqrt{\epsilon}$,} \\
 0 & \text{else,} 
\end{cases}
\end{align}
where the overall mutation rate is given by
\begin{align}
\Ub & = \Lb \int \mu f_0(\mu,s) \, d\mu \, ds 
    = \frac{\Lb \tilde{R} \sqrt{\epsilon}}{4 N c} \int_{\epsilon}^\infty \xi^{-1} e^{-\xi} \, d\xi \, , \nonumber \\
	& \approx \frac{\vo}{2 N c^2} \log \left( \frac{1}{\epsilon} \right) \, .
\end{align}
These expressions show that (within the context of the SSWM limit) the scaling of $\Xavg(t)$ and $\Mavg(t)$ with $N\Ub$ cannot be used to distinguish between the generalized running out of mutations model and the global diminishing returns model in \eq{eq:wiser-epistasis}. However, the time-dependent DFE, $\rho(s,t)$, differs between the two models, which implies that they still constitute different models of macroscopic epistasis. In principle, these differences can be elucidated by considering additional observables or by probing the scaling with $N$ and $\Ub$ in the clonal interference regime.

\section{Comparison with reconstruction data}
\setcounter{equation}{0}

\noindent In the main text, we have focused on signatures of epistasis in long-term patterns of fitness and mutation accumulation. Since our inferences are conducted at this aggregate level, it is natural to ask how our results relate to more traditional measures of epistasis derived from the fitness effects of individual mutations. Actual data in this case is somewhat limited, given the general difficulty of identifying and reconstructing mutations that arose during the course of an evolution experiment. Fortunately, in the case of the LTEE, a small scale study was recently carried out by \citet{khan:etal:2011}, who reconstructed all $2^5$ allelic combinations of the first 5 mutations that fixed in the Ara-1 population. This allowed them to measure the fitness effects of 5 mutations in 16 different genetic backgrounds, 3 of which showed signatures of diminishing returns epistasis.

One the one hand, it is tempting to apply our global diminishing returns models directly to the full dataset, e.g., plotting $s(X)/s(0)$ as a function of $X$ for each of the 5 mutations. \citet{wiser:etal:2013} employed a related method to support the global diminishing returns model in \eq{eq:wiser-epistasis}. However, as we have argued above, we must be careful when extrapolating from macroscopic models of epistasis to fundamentally microscopic measurements. In particular, we showed in the text that one cannot define a consistent model of microscopic epistasis where $s(X) = s(0) e^{-X/X_c}$, since this would violate the bookkeeping property \citep{nagel:etal:2012}. Thus, in a technical sense, the dependence of $s(X)/s(0)$ on $X$ cannot provide additional evidence for the model in \eq{eq:wiser-epistasis}. 

\begin{figure}
\includegraphics{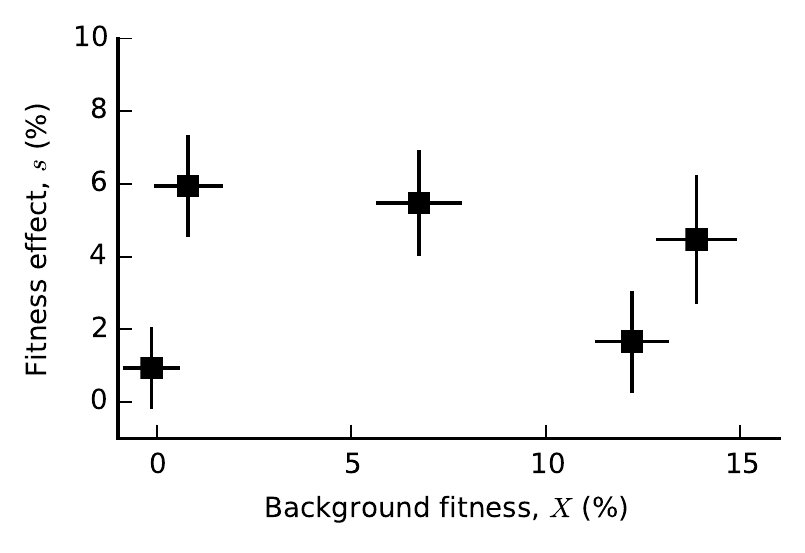}
\caption{The fitness effects (along the line of descent) for the first 5 mutations to fix in the Ara-1 population, estimated from the genetic reconstruction data of \citet{khan:etal:2011} (see Appendix). On the $x$-axis, each mutation is plotted according to the fitness of the genetic background in which it arose. Error bars denote 2 stderr confidence intervals.  \label{fig:khan-trajectory}}
\end{figure}

However, there is one aspect of the reconstruction data that \emph{can} be used to gauge the support for the macroscopic model in \eq{eq:wiser-dfe}. If global diminishing returns is responsible for the decelerating fitness trajectory in \fig{fig:trajectory}, then the fitness effects of fixed mutations should decrease along the line of descent, independent of their effects in other backgrounds. For example, in the SSWM limit, the distribution of fitness effects of fixed mutations (measured in the background in which they arise) must satisfy the scaling relation,
\begin{align}
\rho_f(s|X) \propto \left( \frac{s}{f(X)} \right) \rhoo \left( \frac{s}{f(X)} \right) \, . \label{eq:rhof}
\end{align}
This distribution becomes more complicated in the clonal interference regime, but the scaling with $f(X)$ is approximately preserved up to logarithmic corrections \citep{good:etal:2012}. In either case, we expect the fitness effects of fixed mutations to decrease roughly proportionally to $f(X)$. However, the data itself shows no such decrease (\fig{fig:khan-trajectory}). Of course, this finding should be treated with a degree of caution, since it is based on a sample of five mutations from $\rho_f(s)$ with considerable sampling noise. Nevertheless, it implies that there is limited evidence in the \citet{khan:etal:2011} data to suggest that global diminishing returns is driving the deceleration in the fitness trajectory. As suggested by \citet{draghi:plotkin:2013}, the stronger diminishing returns signal off the line of descent may reflect ascertainment biases inherent in any set of fixed mutations. 

\clearpage

\begin{table}[h!]
\begin{tabular}{|c|c|c|c|}
\hline
Generation & Clone ID & Population & Source \\\hline
2,000 & REL1164A & Ara$-1$ & \citet{barrick:etal:2009} \\\hline
5,000 & REL2179A & Ara$-1$ & \citet{barrick:etal:2009} \\\hline
10,000 & REL4536A &  Ara$-1$ & \citet{barrick:etal:2009} \\\hline
15,000 & REL7177A & Ara$-1$ & \citet{barrick:etal:2009} \\\hline
20,000 & REL8593A & Ara$-1$ & \citet{barrick:etal:2009} \\
	& REL8593A, REL8593B, REL8593C & Ara$-1$ & \citet{wielgoss:etal:2011} \\\hline
30,000 & ZDB16, ZDB357 & Ara$-3$ & \citet{wielgoss:etal:2011} \\\hline
40,000 & REL10947, REL10948, REL10949 & Ara$-5$ & \citet{wielgoss:etal:2011}  \\
 & REL11005, REL11006 & Ara$-6$ & \citet{wielgoss:etal:2011} \\
 & REL11008, REL11009 & Ara$+1$ & \citet{wielgoss:etal:2011} \\
 &  REL10950, REL10951 & Ara$+2$ & \citet{wielgoss:etal:2011} \\
 & REL10956, REL10957 & Ara$+4$ & \citet{wielgoss:etal:2011} \\
 & REL10982, REL10983 & Ara$+5$ & \citet{wielgoss:etal:2011} \\\hline
\end{tabular}
\caption{A list of the clones used to estimate the mutation trajectory in \fig{fig:trajectory}B. \label{table:clone-list}}
\end{table}

\begin{figure}[h!]
\includegraphics{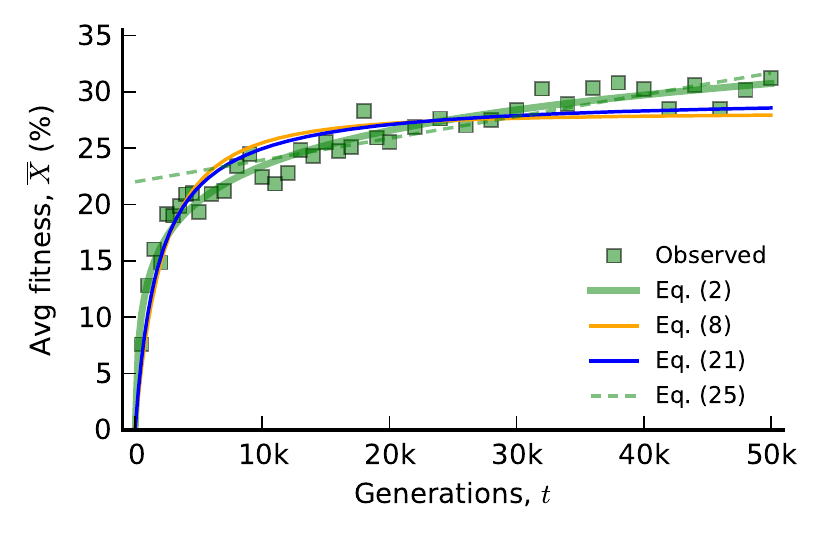}
\caption{Comparison of the four analytical fitness trajectories in Eqs.~(\ref{eq:wiser-trajectory}), (\ref{eq:finite-exponential-trajectory}), (\ref{eq:hyperbolic-trajectory}), and (\ref{eq:nonepistatic-trajectory}). The best-fit parameters were obtained by minimizing the mean squared error, and the fit of the linear trajectory in \eq{eq:nonepistatic-trajectory} was restricted to the last 40,000 generations of evolution.  \label{fig:trajectory-curvefits}}
\end{figure}

\begin{figure}[h!]
\includegraphics{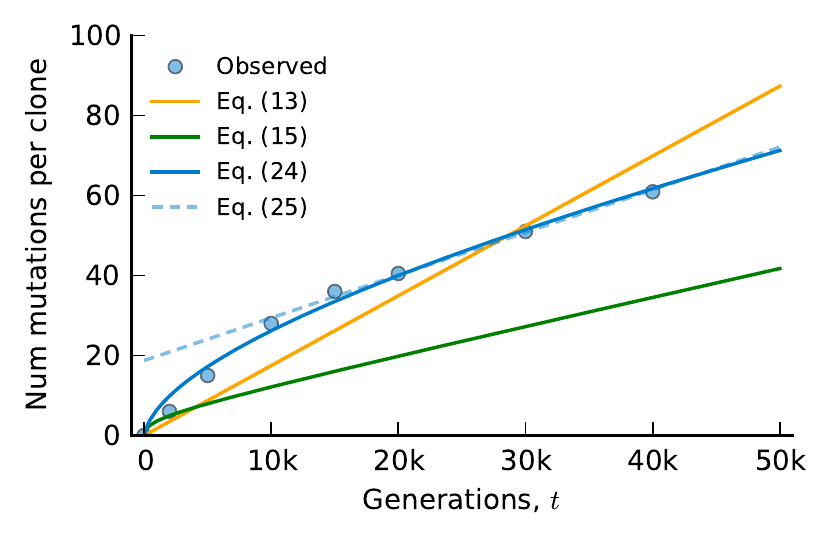}
\caption{Comparison of the four analytical mutation trajectories in Eqs.~(\ref{eq:finite-wiser-mutation-trajectory}), (\ref{eq:uncorrelated-mutation-trajectory}), (\ref{eq:wiser-mutation-trajectory}), and (\ref{eq:nonepistatic-trajectory}). The best-fit parameters were obtained by minimizing the mean squared error in the presence of an unknown neutral mutation rate $0 \leq \Un \leq \Utot$, and the fit of the linear trajectory in \eq{eq:nonepistatic-trajectory} was restricted to the last 40,000 generations of evolution.  \label{fig:mutation-curvefits}}
\end{figure}

\begin{figure}[h!]
\includegraphics{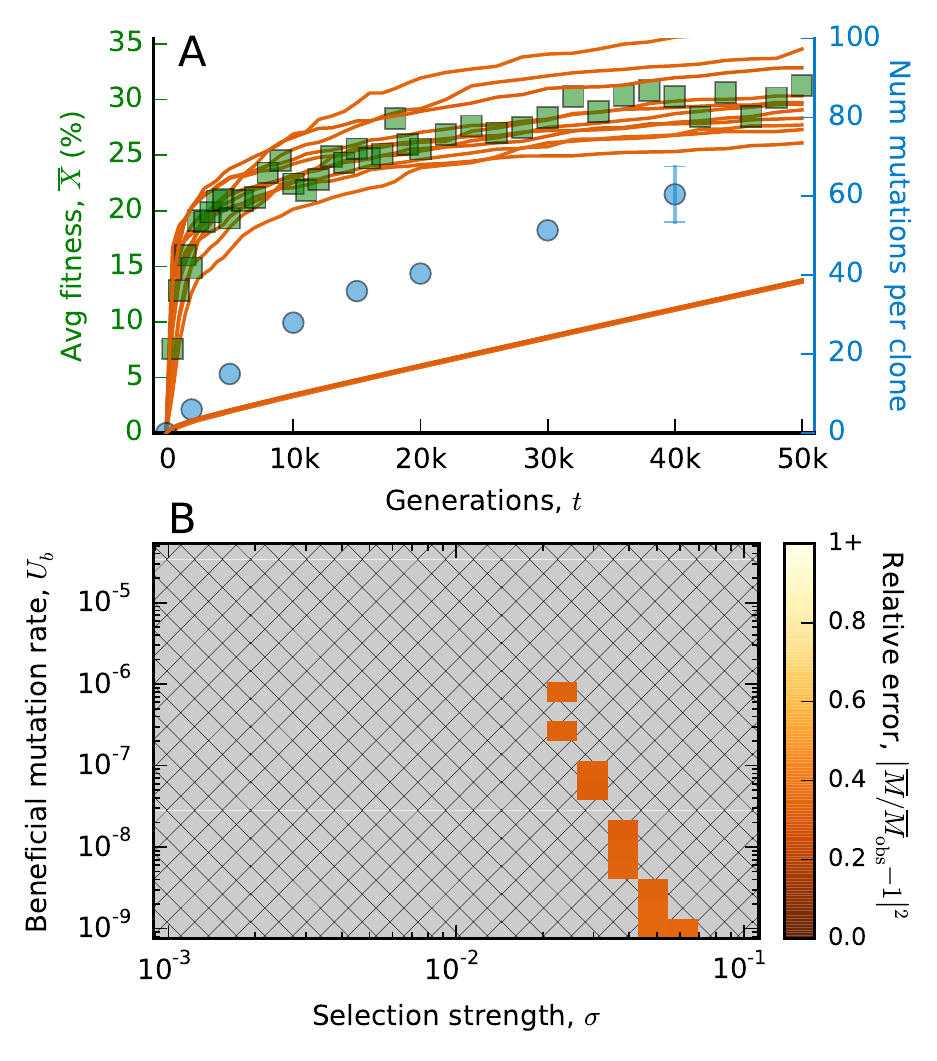}
\caption{Fitting an uncorrelated landscape model to the LTEE data. An analogous version of \fig{fig:finite-sites-fit} constructed for the uncorrelated landscape model in \eq{eq:uncorrelated-epistasis} with an exponential ancestral DFE. Note the change in scale for the relative error in the mutation trajectory. To ensure better convergence of the ensemble mean fitness trajectory, we simulated 100 independent populations for bootstrap resampling instead of the 18 used for the other models. \label{fig:uncorrelated-fit}}
\end{figure}

\begin{figure}[t]
\centering
\includegraphics{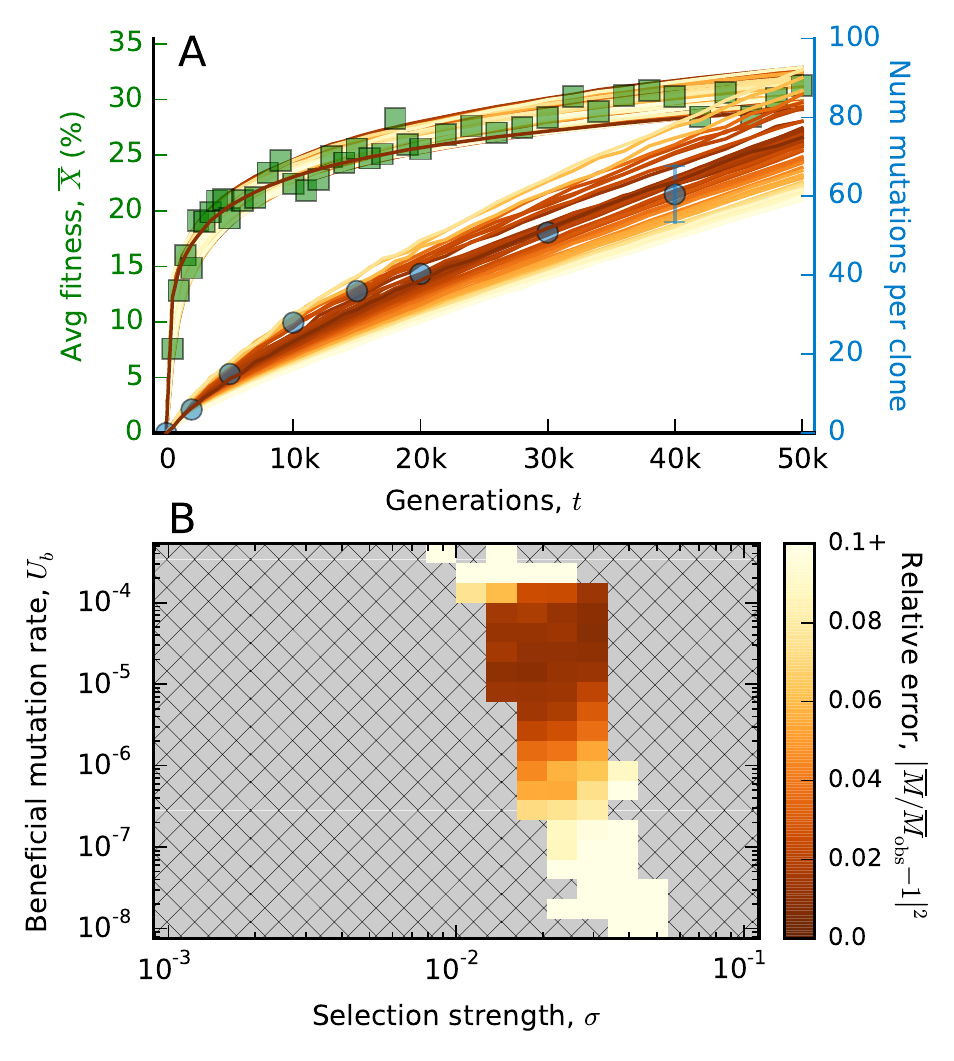} 
\caption{An analogous version of \fig{fig:wiser-fit} for a truncated exponential ancestral DFE ($s_\mathrm{max} = 4\sigma$). \label{fig:modified-wiser-fit}}
\end{figure}

\begin{figure*}
\centering
\includegraphics[width=0.9\textwidth]{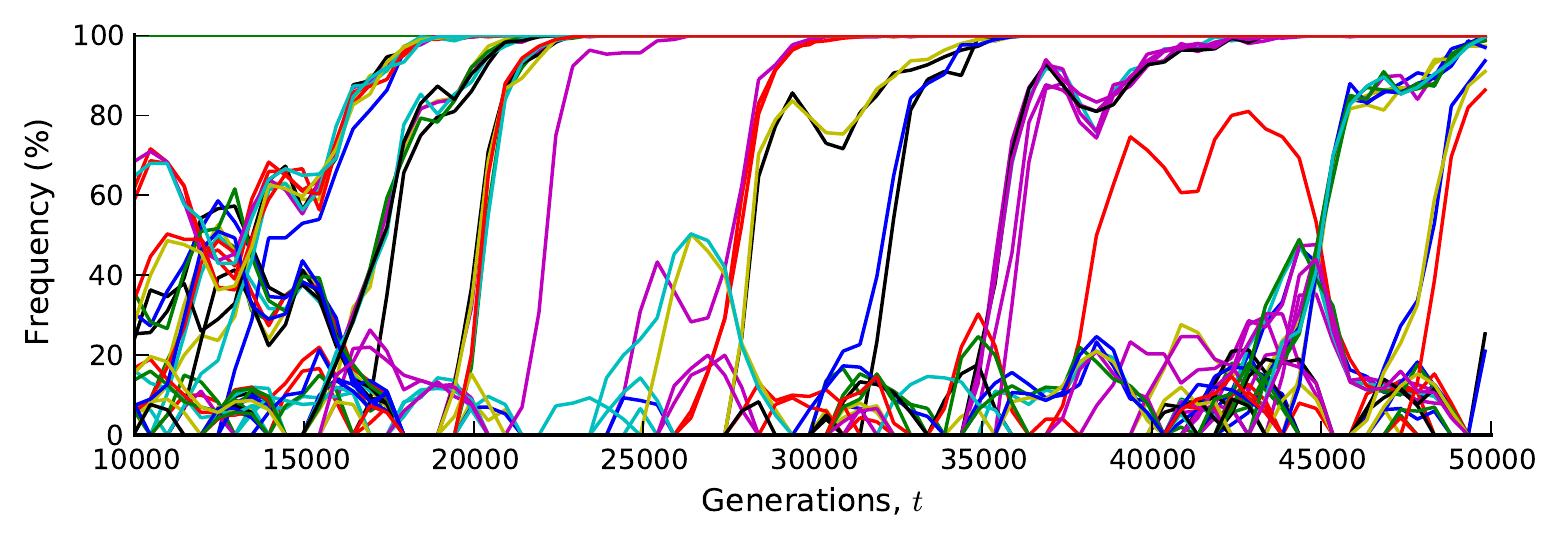} \\
\includegraphics[width=0.9\textwidth]{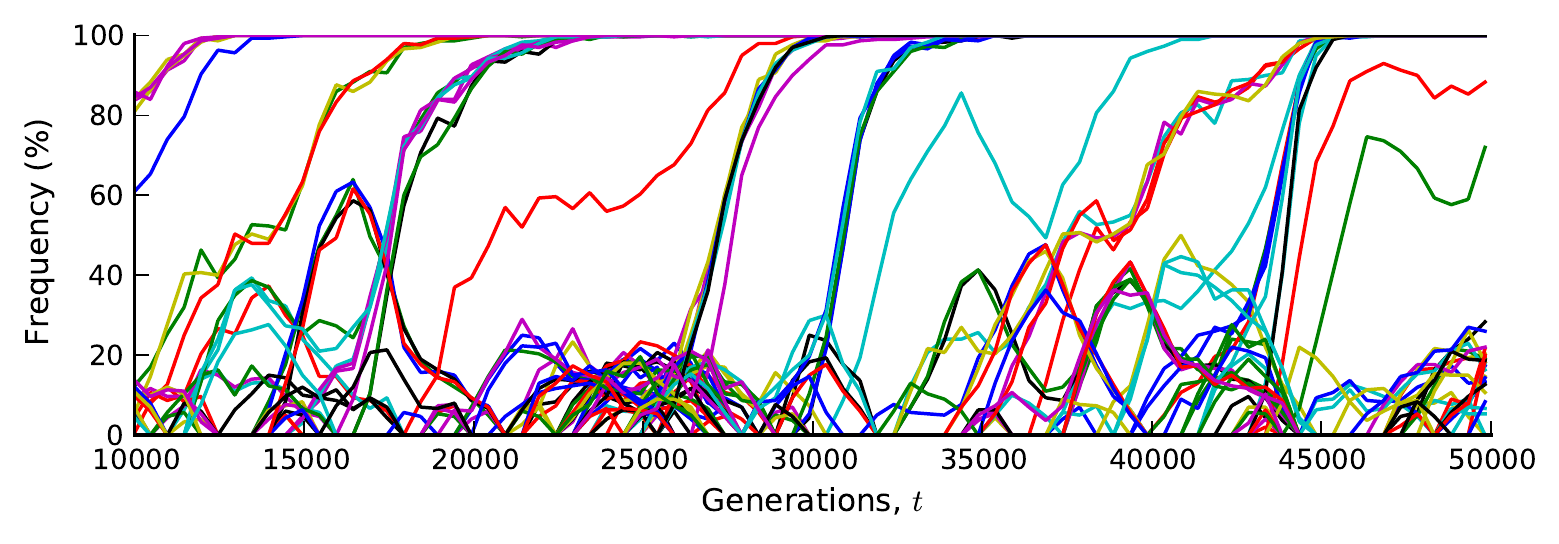}
\caption{Simulated mutational dynamics for the constant truncated exponential DFE in \fig{fig:after-10k-fit}. Colored lines depict the frequencies of all mutations that rose above $5\%$ in the first 50,000 generations. To mimic whole-population sequencing in the LTEE, mutation frequencies are sampled in 500 generation intervals with binomial sampling noise ($n=300$). Parameter values are $\Ub = 2.6 \times 10^{-6}$, $\sigma = 1.1 \times 10^{-3}$, and $\Un = 4 \times 10^{-4}$ (top) and $\Ub = 2.5 \times 10^{-5}$, $\sigma = 7 \times 10^{-4}$, and $\Un = 1.3 \times 10^{-4}$ (bottom). \label{fig:hypothetical-timecourse}}
\end{figure*}

\end{document}